\documentclass[aps,pra,twocolumn]{revtex4-1}

\usepackage[english]{babel}
\usepackage[T1]{fontenc}
\usepackage[utf8]{inputenc}

\usepackage{bbm}  
\usepackage{graphicx}
\usepackage{amsmath}
\usepackage{amssymb}
\usepackage{amsfonts}
\usepackage{braket}
\usepackage{booktabs}
\usepackage{multirow}
\usepackage{simplewick}
\usepackage{bm} 
\usepackage{xcolor}
\usepackage{ulem}
\usepackage{threeparttable}
\usepackage{subfig}
\usepackage{float}
\usepackage{cancel}
\usepackage{algorithm2e}

\usepackage{cellspace}
\newcommand\cincludegraphics[2][]{\raisebox{-0.4\height}{\includegraphics[#1]{#2}}}

\usepackage[autostyle=true]{csquotes}

\graphicspath{{PDF/}}

\definecolor{myred}{rgb}{1 0.7 0.7}
\definecolor{myblue}{rgb}{0.7 0.7 1}
\definecolor{mygreen}{rgb}{0.5 0.9 0.5}
\definecolor{mygray}{gray}{0.8}

\newcommand{\highlight}[2][yellow]{\mathchoice%
  {\colorbox{#1}{$\displaystyle#2$}}%
  {\colorbox{#1}{$\textstyle#2$}}%
  {\colorbox{#1}{$\scriptstyle#2$}}%
  {\colorbox{#1}{$\scriptscriptstyle#2$}}}%

\begin{document}                                                                                                                                                                   
                                                                                                                                                                                   
\title{A correctly scaling rigorously spin-adapted and spin-complete open-shell CCSD implementation for arbitrary high-spin states}
\author{Nils Herrmann}
\email{N.Herrmann@uni-koeln.de}
\author{Michael Hanrath}
\email{Michael.Hanrath@uni-koeln.de}

\affiliation{%
Institute for Theoretical Chemistry, University of Cologne, 
Greinstra\ss e 4, 50939 Cologne, Germany
}%

\date{\today}

\begin{abstract}
In this paper, we report on a correctly scaling novel coupled cluster singles and doubles (CCSD) implementation for arbitrary high-spin open-shell states. The chosen cluster operator is completely spin-free, i.e. employs spatial substitutions only. It is composed of our recently developed L\"owdin-type operators (doi: 10.1063/5.0026762), which ensure (1) spin completeness and (2) spin adaption i.e. spin purity of the CC wave function. In contrast to the proof-of-concept matrix-representation-based implementation presented there, the present implementation relies on second quantization and factorized tensor contractions. 
The generated singles and doubles operators are embedded in an equation generation engine. In the latter, Wick's theorem is used to derive prefactors arising from spin integration directly from the spin-free full contraction patterns. The obtained Wick terms composed of products of Kronecker deltas are represented by special non-antisymmetrized Goldstone diagrams. Identical (redundant) diagrams are identified by solving the underlying graph isomorphism problem.  
All non-redundant graphs are then automatically translated to locally -- one term at a time -- factorized tensor contractions. 
Finally, the spin-adapted and spin-complete (SASC) CCS and CCSD variants are applied to a set of small molecular test systems. Both correlation energies and amplitude norms hint towards a reasonable convergence of the SASC-CCSD method for a BCH series truncation of order four. In comparison to spin orbital CCSD, SASC-CCSD leads to slightly improved correlation energies with differences of up to 0.886\,m$E_H$ (0.98\% w.r.t.\ spin orbital CCSD(T)).

\end{abstract}

\maketitle

\section{Introduction}

The CC method, pioneered by Coester and K\"ummel\cite{Coester1958, Coester1958a}, emerged as one of the most successful ab-initio methods of modern quantum chemistry. Within its framework, the cluster operator $\hat{T}$ is applied to a reference $\ket{\Psi_0}$ to generate the CC wave function $\ket{\Psi_{\text{CC}}}$ via 
\begin{equation*}
\ket{\Psi_{\text{CC}}} = e^{\hat{T}}\ket{\Psi_0}\,.
\end{equation*}

In the most commonly used similarity-transformed CC approach, $\ket{\Psi_0}$ is confined to a single Hartree-Fock determinant while $\hat{T}$ is expressed by spin orbital substitutions moving particles from the occupied to the virtual orbital space. Despite tremendous progress in the last decades\cite{Mukherjee1975, Nakatsuji1978, Nakatsuji1978a, Rittby1988, Yuan2000, Neogrady1992, Neogrady1994, Neogrady1995, Szalay1997, Szalay2000, Berente2002, Heckert2006, Wilke2011, Knowles1993, Knowles2000, Li1993, Li1994, Piecuch1994, Li1994a, Li1995, Li1995a, Li1995b, Jeziorski1995, Jankowski1999, Janssen1991, Nooijen1996, Nooijen2001, Sen2012, Datta2008, Datta2009, Datta2011, Datta2013, Datta2014, Datta2015, Datta2019, Herrmann2020a, Herrmann2021}, the formulation of a general coupled cluster (CC) approach to treat (high-spin) open-shell cases is still an ongoing research field.  In this work, we focus specifically on single restricted open-shell Hartree-Fock (ROHF) type references resembling $S=S_z$ high-spin states. 

Through the application of a usual spin orbital cluster operator $\hat{T}$ (incorporating e.g. spin orbital substitutions) to open-shell references $\ket{\Psi_0}$ (see e.g. \cite{Mukherjee1975}), spin contamination is introduced into the CC wave function even if $\ket{\Psi_0}$ is represented by a genuine spin eigenfunction (see e.g. \cite{Stanton1994, Krylov2000, Li2000}). This contamination arises due to $\hat{T}$  being non-commutative with the $\hat{S}^2$ operator, in general.

Several approaches to adapt the cluster operator $\hat{T}$ (as originally motivated by the symmetry-adapted cluster operators of Nakatsuji and Hirao\cite{Nakatsuji1978, Nakatsuji1978a}) such that it commutes with $\hat{S}^2$ -- effectively removing all spin contamination of the CC wave function -- were proposed in the literature. Naturally, the special case of closed-shell spin adaption -- being fundamentally simpler than the general open-shell case -- received much attention especially in the early years of spin-adapted CC theory (see e.g. \cite{Paldus1977, Adams1979, Ghosh1982, Aadnan1982, Takahashi1985, Jeziorski1988, Piecuch1989, Piecuch1990, Piecuch1992, Piecuch1995, Kondo1995, Kondo1996, Matthews2013, Matthews2015}). 

The present work, however, is concerned with the more general open-shell cases.  
In the literature, projective methods\cite{Rittby1988, Yuan2000} were developed, where spin-contaminated parts of the spin orbital CC expansion are removed applying an appropriate projector. Analogously, restrictive methods\cite{Neogrady1992, Neogrady1994, Neogrady1995, Szalay1997, Szalay2000, Berente2002, Heckert2006, Wilke2011} were proposed explicitly restricting the CC amplitudes such that spin contamination is removed. 
In the latter approaches, Neogr\'{a}dy et al. developed spin-adapted linear CCSD\cite{Neogrady1992}, full CCSD\cite{Neogrady1994} as well as CCSD(T)\cite{Neogrady1995} for the doublet spin state. Szalay and Gauss\cite{Szalay1997} developed a restricted CCSD formalism, which was later expanded to treat excited states\cite{Szalay2000}, to the inclusion of full triples\cite{Berente2002} as well as to an explicitly correlated R12 approach\cite{Wilke2011}. In this formalism, the correct $\hat{S}^2$ expectation value of the CC wave function is enforced by the inclusion of explicit spin equations. To make their approach feasible, the proposed spin equations are followed in a truncated manifold only. The resulting CC wave functions are therefore not spin-adapted in general. 
Later, a comparison of spin-adapted and spin-restricted CCSDs was given by Heckert et al.\cite{Heckert2006}. Furthermore, a partially spin-adapted CCSD was presented by Knowles and Werner\cite{Knowles1993, Knowles2000}, where the cluster operator is composed of a mix of spin-free (spatial) and spin orbital substitutions such that the linear CCSD terms are spin-adapted while the non-linear terms, hence the CC wave function $e^{\hat{T}}\ket{\Psi_0}$, is not.

A different approach to achieve spin adaption is followed by the so-called unitary group approach (UGA)\cite{Li1993, Li1994, Piecuch1994, Li1994a, Li1995, Li1995a, Li1995b, Jeziorski1995, Jankowski1999}. Here, unitary group theory is applied to generate spatial, i.e. spin-free, orbital substitutions such that orthonormal configuration state functions (CSFs) are built. A cluster operator built by these substitutions is guaranteed to produce a properly spin-adapted CC wave function. Through careful choice of the selected substitutions, the completeness of the spanned spin space may also be reached. Linear CCSD\cite{Li1993} and full CCSD\cite{Li1994} theories with special emphasis on doublets, triplets and open-shell singlets were developed and analyzed w.r.t. ROHF doublet\cite{Li1995a} and open-shell singlet and triplet\cite{Li1995b} instabilities. Further applications include the $^1$A$_1$, $^3$B$_1$ separation of the methylene molecule\cite{Piecuch1994, Li1994a}, calculations of the ozone molecule\cite{Li1995} and an analysis of van der Waals interactions of high-spin systems\cite{Jankowski1999}.

As discussed by Matthews et al.\cite{Matthews2013, Matthews2015} for the closed-shell case, theory and implementation of orthogonal UGA CC are quite complex compared to non-orthogonal approaches. In the latter, the cluster operator may be composed of solitary spatial substitution operators instead of linear combinations of such. This not only simplifies equation derivation, but also enables efficient implementations. Already in 1991, Janssen and Schaefer III presented an automated equation derivation engine\cite{Janssen1991}, which was also applied to a non-orthogonal spin-adapted CCSD for high-spin open shell states. Although their approach allows for spectating substitutions, terms required for spin completeness in triplet and higher spin states are missing from their cluster operators\cite{Herrmann2020a}.

Following the early suggestion of Lindgren\cite{Lindgren1978} to approximate the wave operator $e^{\hat{T}}$ by its normal-ordered form $\{e^{\hat{T}}\}$, several CCSD\cite{Nooijen1996, Nooijen2001} and MRCCSD\cite{Sen2012} implementations emerged. While the assumed normal ordering of the wave operator removes a lot of complexity from the theory (mainly by avoiding internal $\hat{T}$ contractions), its effects on the quality of the CC wave function are unknown.  

One of the latest developments in spin-adapted CC theory is the combinatoric open-shell CC (COSCC) approach\cite{Datta2008, Datta2009, Datta2011, Datta2013, Datta2014, Datta2015, Datta2019}. Here, the wave operator is assumed to be normal-ordered while contractions among the cluster operators are reintroduced following a combinatoric cluster expansion. Theories were developed for single-reference CC\cite{Datta2008} as well as state-universal\cite{Datta2009} and state-selective\cite{Datta2011} MRCC. An automated implementation engine was developed\cite{Datta2013} and used to generate COSCCSD equations for doublet spin states. The approach was further expanded to the calculation of analytic first derivatives\cite{Datta2014}, spin densities\cite{Datta2015} and hyperfine coupling tensors\cite{Datta2019}. 

Overall, the published spin-adapted open-shell CC methods seem to be limited by the doubles truncation and the triplet spin state. 

In this work, we present a newly developed CCSD implementation capable of treating arbitrary high-spin states. The presented approach utilizes our previously developed L\"owdin-type non-orthogonal spatial substitution operators\cite{Herrmann2020a} to guarantee (i) spin adaption and (ii) spin completeness of the CC wave function. While (i) is automatically fulfilled by the usage of spin-free substitution operators, (ii) requires special care as to the selection of (usually linear-dependent) substitution operators. In our previous work\cite{Herrmann2020a}, this is ensured by L\"owdin's projection operator method.

The spin-adapted and spin-complete (SASC) CCSD implementation, presented here, utilizes a single set of CC equations. These are applicable to arbitrary $S=S_z$ states without any further equation generation or compiling. In principle, the presented methodology is directly applicable to higher CC truncation levels as well. For the sake of this publication, however, we limit ourselves to the spatial SD manifold (involving spin orbital substitutions of up to rank four). 

In section \ref{Theory}, a short introduction to the SASC-CC theory is given. Section \ref{Methodology} then focuses on the automated equation generation and implementation of the SASC-CCSD method, while section \ref{Application} presents proof-of-concept applications on small test molecules of spin states up to $S=S_z=4$. 


\section{Theory}
\label{Theory}

In this section, theoretical intricacies for the presented CCSD are given. Subsection \ref{SpinAdaption} features a short recap of the non-orthogonal, spin-adapted and spin-complete CC framework -- introduced in our previous work\cite{Herrmann2020a} -- while subsection \ref{SpinfreeCCequations} introduces the spin-free CCSD equations considered in this work.

\subsection{Spin adaption and spin completeness in the CC framework}
\label{SpinAdaption}

As already introduced in \cite{Herrmann2020a, Herrmann2021}, we denote doubly occupied spatial orbitals contained in $\mathbb{O}$ by $i,j,\ldots$, virtual spatial orbitals contained in $\mathbb{V}$ by $a, b, \ldots$ and singly occupied active orbitals contained in $\mathbb{A}$ by $v, w, \ldots$. Arbitrary $\nu$-fold spatial substitutions in the high-spin case are given by 
\begin{equation}
\hat{E}_{p_1\ldots p_{\nu}}^{q_1\ldots q_{\nu}} = \sum_{\sigma_1 = \alpha , \beta} \ldots \sum_{\sigma_{\nu} = \alpha , \beta} \hat{X}_{p_1\sigma_1 \ldots p_{\nu}\sigma_{\nu}}^{q_1\sigma_1\ldots q_{\nu}\sigma_{\nu}}\,,
\end{equation}

where $p_1\ldots p_{\nu} \in \mathbb{O}\cup\mathbb{A}$ and $q_1\ldots q_{\nu} \in \mathbb{A}\cup\mathbb{V}$. The $\hat{X}$ operators denote spin orbital substitution operators given by their usual second quantized form as
\begin{equation}
\hat{X}_{p_1\sigma_1 \ldots p_{\nu}\sigma_{\nu}}^{q_1\sigma_1\ldots q_{\nu}\sigma_{\nu}} = \hat{a}^{\dagger}_{q_1\sigma_1} \ldots \hat{a}^{\dagger}_{q_{\nu}\sigma_{\nu}}\hat{a}_{p_{\nu}\sigma_{\nu}}\ldots\hat{a}_{p_1\sigma_1}\,,
\end{equation}

\newpage
\onecolumngrid
\begin{center}
\begin{align}
\label{Singles.eq}
\hat{T}_1 &= 
\sum_{i,a} \left( t_i^a \hat{E}_i^a + \sum_v t_{vi}^{av} \hat{E}_{vi}^{av} \right) 
+ \sum_{i,v} t_i^v \hat{E}_i^v 
+ \sum_{v,a} t_v^a \hat{E}_v^a \\
\label{Doubles.eq}
\hat{T}_2 &= 
\sum_{i,a} \left( t_{ii}^{aa} \hat{E}_{ii}^{aa}
+ \sum_{v} \left[ t_{ii}^{av} \hat{E}_{ii}^{av}
+ t_{iv}^{aa} \hat{E}_{iv}^{aa} \right]
+ \sum_{v\neq w} t_{iw}^{va} \hat{E}_{iw}^{va} \right)
+ \sum_{i,a<b} \left( t_{ii}^{ab} \hat{E}_{ii}^{ab}
+ \sum_{v} \left[ t_{iv}^{ab} \hat{E}_{iv}^{ab} 
+ t_{vi}^{ab} \hat{E}_{vi}^{ab}
+ t_{vii}^{abv} \hat{E}_{vii}^{abv} \right] \right.
\\ \notag &\phantom{=}
+ \left. \sum_{v\neq w} t_{wvi}^{abw} \hat{E}_{wvi}^{abw} \right)
+ \sum_{i<j,a} \left( t_{ij}^{aa} \hat{E}_{ij}^{aa}
+ \sum_{v} \left[ t_{ij}^{av} \hat{E}_{ij}^{av}
+ t_{ji}^{av} \hat{E}_{ji}^{av}
+ t_{vji}^{aav} \hat{E}_{vji}^{aav} \right]
+ \sum_{v\neq w} t_{wji}^{avw} \hat{E}_{wji}^{avw} \right)
+ \sum_{i<j, a<b}  \left( \rule{0pt}{0.7cm} t_{ij}^{ab} \hat{E}_{ij}^{ab} \right. 
\\ \notag &\phantom{=} \left.
+ t_{ji}^{ab} \hat{E}_{ji}^{ab}
+ \sum_{v} \left[ t_{vij}^{abv} \hat{E}_{vij}^{abv}
+ t_{ivj}^{abv} \hat{E}_{ivj}^{abv}
+ t_{vji}^{abv} \hat{E}_{vji}^{abv} \right]
+ \sum_{v<w} t_{vwij}^{abvw} \hat{E}_{vwij}^{abvw} \right)
+ \sum_{i<j, v<w} t_{ij}^{vw} \hat{E}_{ij}^{vw}
+ \sum_{v<w, a<b} t_{vw}^{ab} \hat{E}_{vw}^{ab}
\end{align}
\end{center}
\twocolumngrid

where $\sigma_1\ldots\sigma_{\nu}$ denote spin variables assigned to spatial orbitals $1\ldots\nu$.

The spin-free $\hat{E}$ operators apply pair-wise spin-summed substitutions such that the original spin state of their target remains unaltered. Any cluster operator $\hat{T}$ composed solely of $\hat{E}$ operators is therefore properly spin-adapted, i.e. does not produce spin-contaminated wave functions (provided the reference determinant does not include spin contamination). However, depending on the number of open shells and the total spin quantum number $S$, the $\hat{T}$ operator may span an insufficient spin space. For a more detailed definition of spin adaption and spin completeness, please refer to our previous work\cite{Herrmann2020a}.

Special care needs to be taken to construct cluster operators, which (i) are spin-adapted, (ii) produce complete spin spaces for arbitrary spin states and (iii) do not involve linear-dependent substitutions. Such operators were generated in \cite{Herrmann2020a} by the use of L\"owdin's projection operator method. These L\"owdin-type operators for the single and double substitutions are given in equations (\ref{Singles.eq}) and (\ref{Doubles.eq}).

\subsection{Spin-free CC equations}
\label{SpinfreeCCequations}

In this work, we aim at solving the similarity transformed CCSD equations with 
\begin{equation}
\label{CCEquations.eq}
\braket{\Psi_x | e^{-\hat{T}} \hat{H}_N e^{\hat{T}} \Psi_0} = \delta_{x0} E_{\text{corr}}\,,
\end{equation}
for $\hat{T} = \hat{T}_1 + \hat{T}_2$ as defined in equations (\ref{Singles.eq}) and (\ref{Doubles.eq}) and $\hat{H}_N$ denoting the normal-ordered Hamiltonian. The reference $\ket{\Psi_0}$ is assumed to be a single high-spin ($S = S_z$) ROHF determinant such that 
\begin{equation}
\label{Unmixed.eq}
\braket{\Psi_0 | \hat{T} \Psi_0} = 0
\end{equation}
holds. Since the Hamiltonian $\hat{H}$ commutes with both the $\hat{S}^2$ and the $\hat{S}_z$ operators, it may be defined by spin-free $\hat{E}$ operators:
\begin{equation}
\hat{H} = \sum_{p, q} \underbrace{\braket{p|\hat{h}(1)|q}}_{o_{q}^{p}} \hat{E}_{q}^{p} + \frac{1}{2} \sum_{p, q, r, s} \underbrace{\braket{pq|\hat{g}(1,2)|rs}}_{o^{pq}_{rs}} \hat{E}_{rs}^{pq} 
\end{equation}
Here, the indices $p, q, r$ and $s$ denote arbitrary spatial orbitals from orbital spaces $\mathbb{O}$, $\mathbb{A}$ or $\mathbb{V}$.
In order to apply Wick's theorem to $\hat{H}$, the spin-free $\hat{E}$ operators may be expanded to their spin orbital analogue. Within the usual particle-hole formalism for the high-spin ($S=S_z$) reference $\ket{\Psi_0}$, it is 
\begin{align}
\label{wick_1.eq}
\contraction{}{\hat{b}}{_{p\sigma}^{\dagger}}{\hat{b}}
\hat{b}_{p\sigma}^{\dagger}\hat{b}_{q\omega} &= 0 \\
\contraction{}{\hat{b}}{_{p\sigma}}{\hat{b}}
\hat{b}_{p\sigma}\hat{b}_{q\omega}^{\dagger} &= \left\{\rule{0pt}{1.8cm}\right.
\begin{array}{rc}
p,q\in\mathbb{O}: & \contraction{}{\hat{a}}{_{i\sigma}^{\dagger}}{\hat{a}}\hat{a}_{i\sigma}^{\dagger}\hat{a}_{j\omega} = \delta_{ij}\delta_{\sigma\omega}\\[4pt] 
p,q\in\mathbb{A}: & \left\{\rule{0pt}{0.8cm}\right.
\begin{array}{c}
\contraction{}{\hat{a}}{_{v\alpha}^{\dagger}}{\hat{a}}\hat{a}_{v\alpha}^{\dagger}\hat{a}_{w\alpha} = \delta_{vw} \\[4pt] 
\contraction{}{\hat{a}}{_{v\beta}}{\hat{a}}\hat{a}_{v\beta}\hat{a}_{w\beta}^{\dagger} = \delta_{vw}
\end{array}\vspace{4pt} \\
p,q\in\mathbb{V}: & \contraction{}{\hat{a}}{_{a\sigma}}{\hat{a}}\hat{a}_{a\sigma}\hat{a}_{b\omega}^{\dagger} = \delta_{ab}\delta_{\sigma\omega}  \\[4pt]
\text{else:} & 0
\label{wick_2.eq}
\end{array}
\end{align} 
where $\hat{b}^{\dagger}$ and $\hat{b}$ denote quasi particle/hole creation and annihilation operators, respectively. Clearly, the usual contraction rules (including the spin restriction $\delta_{\sigma\omega}$) hold in the occupied and virtual orbital spaces $\mathbb{O}$ and $\mathbb{V}$. For the singly occupied active space $\mathbb{A}$, the surviving contractions depend on the spin information of the contracted orbitals. For $\alpha$ orbitals (occupied in the spin orbital picture of the high-spin reference), only creator -- annihilator contractions survive, while for $\beta$ orbitals, the opposite is true. This is of particular importance when evaluating spin summations. While contractions in $\mathbb{O}$ or $\mathbb{V}$ leave at least one spin summation intact, contractions in $\mathbb{A}$ break down to either the $\alpha$ or the $\beta$ part of the summation. 

The operators $\hat{E}_q^p$ and $\hat{E}_{rs}^{pq}$ may now be normal-ordered with respect to the given high-spin reference determinant $\ket{\Psi_0}$ (indicated by curly brackets) to yield 
\begin{align} 
\notag
\hat{E}_q^p &= \left\{ \hat{E}_q^p \right\} 
+ \sum_{\sigma = \alpha, \beta} \left\{
\contraction{}{\hat{a}}{_{p\sigma}^{\dagger}}{\hat{a}}
\hat{a}_{p\sigma}^{\dagger}\hat{a}_{q\sigma} \right\} \\ 
&= \left\{ \hat{E}_q^p \right\} + 2\delta_{pq\in\mathbb{O}} + \delta_{pq\in\mathbb{A}} 
\label{Eqp.eq}
\end{align}
and 
\begin{align}
\notag
\hat{E}_{rs}^{pq} &= \left\{\hat{E}_{rs}^{pq}\right\} + \sum_{\sigma_1, \sigma_2 = \alpha, \beta} \left[ 
\left\{\contraction{}{\hat{a}}{_{p\sigma_1}^{\dagger} \hat{a}_{q\sigma_2}^{\dagger} }{\hat{a}} \hat{a}_{p\sigma_1}^{\dagger} \hat{a}_{q\sigma_2}^{\dagger} \hat{a}_{s\sigma_2} \hat{a}_{r\sigma_1}\right\} \right. \\ \notag
&\phantom{=} \left.  
+ \left\{\contraction{}{\hat{a}}{_{p\sigma_1}^{\dagger} \hat{a}_{q\sigma_2}^{\dagger} \hat{a}_{s\sigma_2} }{\hat{a}} \hat{a}_{p\sigma_1}^{\dagger} \hat{a}_{q\sigma_2}^{\dagger} \hat{a}_{s\sigma_2} \hat{a}_{r\sigma_1}\right\}
+ \left\{\contraction{\hat{a}{}_{p\sigma_1}^{\dagger} }{\hat{a}}{_{q\sigma_2}^{\dagger} }{\hat{a}} \hat{a}_{p\sigma_1}^{\dagger} \hat{a}_{q\sigma_2}^{\dagger} \hat{a}_{s\sigma_2} \hat{a}_{r\sigma_1}\right\} \right. \\ \notag
&\phantom{=} \left.  
+ \left\{\contraction{\hat{a}{}_{p\sigma_1}^{\dagger} }{\hat{a}}{_{q\sigma_2}^{\dagger} \hat{a}_{s\sigma_2} }{\hat{a}} \hat{a}_{p\sigma_1}^{\dagger} \hat{a}_{q\sigma_2}^{\dagger} \hat{a}_{s\sigma_2} \hat{a}_{r\sigma_1}\right\} 
+ \left\{
\contraction[1ex]{}{\hat{a}}{_{p\sigma_1}^{\dagger} \hat{a}_{q\sigma_2}^{\dagger} }{\hat{a}}
\contraction[2ex]{\hat{a}_{p\sigma_1}^{\dagger} }{\hat{a}}{_{q\sigma_2}^{\dagger} \hat{a}_{s\sigma_2}}{\hat{a}}
\hat{a}_{p\sigma_1}^{\dagger} \hat{a}_{q\sigma_2}^{\dagger} \hat{a}_{s\sigma_2} \hat{a}_{r\sigma_1}
\right\} \right. \\ \notag
&\phantom{=} \left. 
+ \left\{
\contraction[1ex]{\hat{a}_{p\sigma_1}^{\dagger}}{\hat{a}}{_{q\sigma_2}^{\dagger}}{\hat{a}}
\contraction[2ex]{}{\hat{a}}{_{p\sigma_1}^{\dagger} \hat{a}_{q\sigma_2}^{\dagger} \hat{a}_{s\sigma_2}}{\hat{a}}
\hat{a}_{p\sigma_1}^{\dagger} \hat{a}_{q\sigma_2}^{\dagger} \hat{a}_{s\sigma_2} \hat{a}_{r\sigma_1}
\right\}
\right] \\ \label{Erspq.eq}
&= \left\{\hat{E}_{rs}^{pq}\right\} 
\textcolor{red}{- \delta_{ps\in\mathbb{O}}\left\{\hat{E}_r^q\right\}} 
\textcolor{blue}{- \delta_{ps\in\mathbb{A}}\left\{\hat{X}_{r\alpha}^{q\alpha}\right\}} \\ \notag
&\phantom{=}
+ \left( 2\delta_{pr\in\mathbb{O}} + \delta_{pr\in\mathbb{A}} \right) \left\{\hat{E}_s^q\right\}
+ \left( 2\delta_{qs\in\mathbb{O}} + \delta_{qs\in\mathbb{A}} \right) \left\{\hat{E}_r^p\right\} \\ \notag
&\phantom{=}
\textcolor{red}{- \delta_{qr\in\mathbb{O}} \left\{\hat{E}_s^p\right\}}
\textcolor{blue}{- \delta_{qr\in\mathbb{A}} \left\{\hat{X}_{s\alpha}^{p\alpha}\right\}} - 2\delta_{ps\in\mathbb{O}}\delta_{qr\in\mathbb{O}} \\ \notag
&\phantom{=} 
- \delta_{ps\in\mathbb{O}}\delta_{qr\in\mathbb{A}} - \delta_{ps\in\mathbb{A}}\delta_{qr\in\mathbb{O}} - \delta_{ps\in\mathbb{A}}\delta_{qr\in\mathbb{A}} \\ \notag
&\phantom{=}
+ 4\delta_{pr\in\mathbb{O}}\delta_{qs\mathbb{O}} + 2\delta_{pr\in\mathbb{O}}\delta_{qs\mathbb{A}} + 2\delta_{pr\in\mathbb{A}}\delta_{qs\mathbb{O}} \\ \notag
&\phantom{=}
+ \delta_{pr\in\mathbb{A}}\delta_{qs\mathbb{A}}\,.
\end{align}
 
Interestingly, the single contractions of $\sigma_1$ and $\sigma_2$ spins occurring in $\hat{E}_{rs}^{pq}$ lead to a single spin summation for occupied indices (colored in red) and to a completely broken spin summation for active indices (colored in blue). In the latter case, this leads to pure spin orbital substitutions $\hat{X}$ of $\alpha$ electrons. 

Therefore, $\hat{H}$ may be decomposed into its normal-ordered part $\hat{H}_N$ such that two separate one-particle interrelating operators $\hat{F}_N^{(1)}$ and $\hat{F}_N^{(2)}$ emerge:
\begin{align*} 
\hat{H} &= \underbrace{\hat{F}_N^{(1)} + \hat{F}_N^{(2)} + \hat{V}_N}_{\hat{H}_N} + E_{\text{ROHF}} \\ 
\hat{F}_N^{(1)} &= \sum_{pq}\underbrace{\left[o_q^p + \sum_i \left( 2o_{qi}^{pi} - o_{iq}^{pi}\right) + \sum_v o_{qv}^{pv}\right]}_{f_q^p}  \left\{\hat{E}_q^p\right\} \\ 
\hat{F}_N^{(2)} &= - \sum_{pqv}o_{vq}^{pv}\left\{\hat{X}_{q\alpha}^{p\alpha}\right\} \\ 
\hat{V}_N &= \frac{1}{2} \sum_{pqrs}o_{rs}^{pq}\left\{\hat{E}_{rs}^{pq}\right\}
\end{align*}
For the fully-contracted terms of equations (\ref{Eqp.eq}) and (\ref{Erspq.eq}), the scalar quantity $E_{\text{ROHF}}$ with
\begin{align*}
E_{\text{ROHF}} &= 2\sum_i o_i^i + \sum_v o_v^v + \sum_{ij}\left( 2o_{ij}^{ij} - o_{ji}^{ij} \right) \\ 
&\phantom{=} + \sum_{iv}\left(2o_{iv}^{iv} - o_{vi}^{iv}\right) + \frac{1}{2}\sum_{vw}\left(o_{vw}^{vw} - o_{wv}^{vw}\right)
\end{align*}
emerges, which corresponds to the restricted open-shell Hartree-Fock (ROHF) energy for a single high-spin determinant (c.f. \cite{Roothaan1960}).

\section{Methodology}
\label{Methodology}

In this section, the methodology of the presented CCSD equation derivation and implementation engine is presented. In particular, differences to spin orbital CCSD implementations are highlighted. Figure \ref{GeneralStructure.fig} features a flowchart diagram showing the consecutive steps from BCH expansion to the final factorized code for one particular example. The specifics steps are explained in greater detail throughout subsections \ref{Wick} to \ref{FinalEquations}:

\begin{figure}
\begin{center}
\includegraphics[width=.5\textwidth]{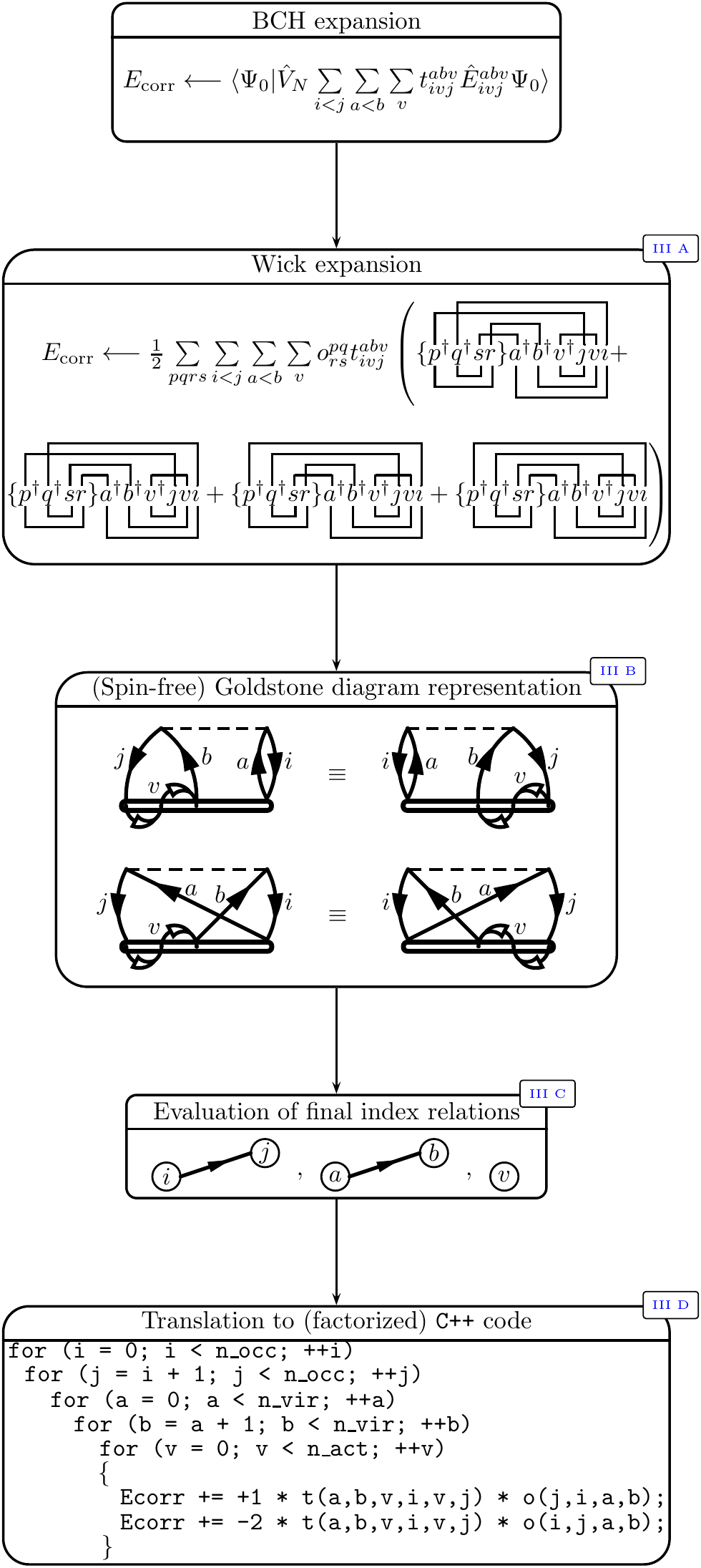}
\caption{\label{GeneralStructure.fig} Flowchart diagram highlighting the consecutive steps from BCH expansion to the final factorized code for one exemplary contribution to the CCSD correlation energy.}
\end{center}
\end{figure}

In subsection \ref{Wick}, the application of Wick's theorem\cite{Wick1950} to the spin-free BCH terms is demonstrated. Connecting lines below the second quantized operator strings are used to abbreviate spin summations. The prefactors arising from spin summations are calculated directly from the full contraction patterns.

After obtaining Wick terms composed of products of Kronecker deltas, summarizable (redundant) terms need to be collected. In this work, we apply special non-antisymmetrized Goldstone diagrams (defined throughout subsection \ref{Goldstone}) to achieve this task. These diagrams are defined such that summarizable terms are represented by isomorphic, i.e. topologically equivalent, graphs. 
 
Finally, diagram-wise index relations need to be evaluated for each Goldstone topology (c.f. subsection \ref{IndexRelations}). In this work, we introduce graphs -- denoted index relation graphs -- to evaluate, keep track and analyze these index relations. 
 
In subsection \ref{FinalEquations}, the generation of the final factorized \texttt{C++} code is outlined. Factorization routes are optimized locally, i.e. diagram-wise, by a complexity estimation using index relation graphs.

\subsection{Derivation of Wick terms}
\label{Wick}

Following the Baker-Campbell-Hausdorff (BCH) expansion of the CC equations (\ref{CCEquations.eq}), sums of products of second quantized creators and annihilators are obtained. When applying Wick's theorem\cite{Wick1950} to the operator strings, only fully contracted terms survive. All surviving and vanishing contraction types for the spin orbital CC as well as the spin-adapted and spin-complete CC framework (followed in this work) are shown in Table \ref{SurvivingAndVanishing.tab}.

In the spin orbital CC formulation, the cluster operator $\hat{T}$ is automatically normal-ordered w.r.t. the Fermi vacuum $\ket{\Psi_0}$ such that no internal contractions within the same $\hat{T}$ are possible ($\contraction{\hspace{1pt}}{\rule{0pt}{8pt}}{\hspace{1pt}}{\hat{T}}\hat{T} = 0$). Furthermore, the usual spin orbital $\hat{T}$ includes hole and particle creators -- $\hat{a}_{i\sigma}$ and $\hat{a}_{a\sigma}^{\dagger}$ -- only. For spin orbital CC, this results in no surviving $\contraction{}{\hat{T}}{}{\hat{T}}\hat{T}\hat{T}$ and $\contraction{}{\hat{T}}{}{\hat{H}}\hat{T}\hat{H}_N$ contractions, which leads to a drastically simplified BCH expansion. 

Unfortunately, none of these simplification hold true in the general spin-adapted and spin-complete case. Due to the presence of active indices $v, w, \ldots$ corresponding to either occupied or virtual spin orbitals in the high-spin 

\begin{center}
\begin{table}[h]
\caption{\label{SurvivingAndVanishing.tab}Surviving and vanishing contraction types in the spin orbital framework (implying $\mathbb{O}\cap\mathbb{V} = \emptyset$) and in the spin-adapted and spin-complete framework followed in this work.}
\begin{tabular}{c|c|c}
\textbf{Contraction} & \textbf{Spin orbital} & \textbf{Spin-adapted} \\ \hline
$\rule{0pt}{15.5pt}\contraction{\hspace{1pt}}{\rule{0pt}{8pt}}{\hspace{1pt}}{\hat{T}}\hat{T}$ & $=0$ & $\neq 0$ \\ \hline
$\rule{0pt}{16pt}\contraction{}{\hat{T}}{}{\hat{T}}\hat{T}\hat{T}$ & $=0$ & $\neq 0$ \\ \hline
$\rule{0pt}{16pt}\contraction{}{\hat{T}}{}{\hat{H}}\hat{T}\hat{H}_N$ & $=0$ & $\neq 0$ \\ \hline 
$\rule{0pt}{16pt}\contraction{}{\hat{H}}{_N}{\hat{T}}\hat{H}_N\hat{T}$ & $\neq 0$ & $\neq 0$
\end{tabular}
\end{table}
\end{center}

reference, some operators in $\hat{T}$ may contract from the left. Additionally, spectating indices, as e.g. in $\hat{E}_{vi}^{av}$, lead to non-vanishing internal contractions. Despite these general considerations, Wick's theorem may be applied as in the spin orbital case. However, special care needs to be taken when evaluating spin summations. 

Consider, e.g., the following term contributing to $E_{\text{corr}}$: 
\begin{align}
\label{Wickexample.eq}
E_{\text{corr}} \longleftarrow &
\Braket{\Psi_0|\frac{1}{2}\sum_{pqrs}o_{rs}^{pq}\left\{\hat{E}_{rs}^{pq} \right\} \sum_{i<j, a<b} t_{ij}^{ab} \hat{E}_{ij}^{ab}\Psi_0} \\ \notag
&= \frac{1}{2}\sum_{pqrs}\sum_{i<j, a<b} o_{rs}^{pq}t_{ij}^{ab}\sum_{\sigma_1, \sigma_2, \sigma_3, \sigma_4 = \alpha, \beta} \\ \notag
&\phantom{=} \braket{\Psi_0|\left\{\hat{a}_{p\sigma_1}^{\dagger}\hat{a}_{q\sigma_2}^{\dagger}\hat{a}_{s\sigma_2}\hat{a}_{r\sigma_1}\right\} \hat{a}_{a\sigma_3}^{\dagger}\hat{a}_{b\sigma_4}^{\dagger}\hat{a}_{j\sigma_4}\hat{a}_{i\sigma_3}\Psi_0}
\end{align}

From now on, we abbreviate second quantized operators by neglecting the $\hat{a}$ such that
\begin{equation*}
\hat{a}_{p\sigma}^\dagger = (p\sigma)^\dagger \quad \text{and} \quad \hat{a}_{q\sigma} = (q\sigma)
\end{equation*}
and introduce a special notation for spin summations:

\begin{equation*}
\sum_{\sigma = \alpha, \beta} (p \sigma)^\dagger (q\sigma) = 
\bcontraction{}{p}{^\dagger}{q} p^\dagger q\,.
\end{equation*}
Each spatial index pair ($p,q$) summed over the same spin variable ($\sigma$) will be represented by a connecting line below the operator expression, which we will call \textit{spin path}. In contrast to spin paths, Wick contractions will be displayed above the operator strings. 

Using this new notation, the rhs of equation (\ref{Wickexample.eq}) now reads 
\begin{equation}
\frac{1}{2}\sum_{pqrs}\sum_{i<j, a<b} o_{rs}^{pq} t_{ij}^{ab} \left(\left\{
\bcontraction[1ex]{p^{\dagger}}{q}{^{\dagger}}{s}
\bcontraction[2ex]{}{p}{^{\dagger}q^{\dagger}s}{r}
p^{\dagger}q^{\dagger}sr\right\} 
\bcontraction[1ex]{a^{\dagger}}{b}{^{\dagger}}{j}
\bcontraction[2ex]{}{a}{^{\dagger}b^{\dagger}j}{i}
a^{\dagger}b^{\dagger}ji \right)_{FC}\,,
\end{equation}
where the notation $\left(\ldots\right)_{FC}$ shall denote the sum of all full contraction patterns. The expression now contains four unique spin paths containing the pairs $pr$, $qs$, $ai$ and $bj$. Each operator (Wick) contraction -- denoted as contraction lines above the operator strings -- connecting two distinct spin paths adds a Kronecker delta for the participating spin variables such that the two spin summations are broken down to one. In other words, the spin paths of contracted operators merge. By this spin path tracing, the individual prefactors of arbitrary contraction patterns with regard to spin summation are easily derived. Every connected spin path (representing an isolated spin summation over $\alpha$ and $\beta$) in a full contraction pattern, contributes a prefactor of $2$. For the given example, it is 
\begin{align*}
\left(\left\{\hat{E}_{rs}^{pq}\right\}\hat{E}_{ij}^{ab}\right)_{FC} &=  
\textcolor{red}{\bcontraction[1ex]{\{p^\dagger}{q}{^\dagger}{s}}
\textcolor{blue}{\bcontraction[2ex]{\{}{p}{^\dagger q^\dagger s}{r}}
\textcolor{blue}{\bcontraction[1ex]{\{p^\dagger q^\dagger sr\} a^\dagger}{b}{^\dagger}{j}}
\textcolor{red}{\bcontraction[2ex]{\{p^\dagger q^\dagger sr\}}{a}{^\dagger b^\dagger j}{i}}
\textcolor{red}{\contraction[1ex]{\{p^\dagger q^\dagger}{s}{r\} }{a}}
\textcolor{blue}{\contraction[2ex]{\{p^\dagger q^\dagger s}{r}{\} a^\dagger}{b}}
\textcolor{blue}{\contraction[3ex]{\{}{p}{^\dagger q^\dagger sr\} a^\dagger b^\dagger}{j}}
\textcolor{red}{\contraction[4ex]{\{p^\dagger}{q}{^\dagger sr\} a^\dagger b^\dagger j}{i}}
\{p^\dagger q^\dagger sr\} a^\dagger b^\dagger ji
+
\textcolor{red}{\bcontraction[1ex]{\{p^\dagger}{q}{^\dagger}{s}}
\textcolor{red}{\bcontraction[2ex]{\{}{p}{^\dagger q^\dagger s}{r}}
\textcolor{red}{\bcontraction[1ex]{\{p^\dagger q^\dagger sr\} a^\dagger}{b}{^\dagger}{j}}
\textcolor{red}{\bcontraction[2ex]{\{p^\dagger q^\dagger sr\}}{a}{^\dagger b^\dagger j}{i}}
\textcolor{red}{\contraction[1ex]{\{p^\dagger q^\dagger s}{r}{\}}{a}}
\textcolor{red}{\contraction[2ex]{\{p^\dagger q^\dagger}{s}{r\} a^\dagger}{b}}
\textcolor{red}{\contraction[3ex]{\{}{p}{^\dagger q^\dagger sr\} a^\dagger b^\dagger}{j}}
\textcolor{red}{\contraction[4ex]{\{p^\dagger}{q}{^\dagger sr\} a^\dagger b^\dagger j}{i}}
\{p^\dagger q^\dagger sr\} a^\dagger b^\dagger ji \\ 
&\phantom{=} 
+
\textcolor{red}{\bcontraction[1ex]{\{p^\dagger}{q}{^\dagger}{s}}
\textcolor{red}{\bcontraction[2ex]{\{}{p}{^\dagger q^\dagger s}{r}}
\textcolor{red}{\bcontraction[1ex]{\{p^\dagger q^\dagger sr\} a^\dagger}{b}{^\dagger}{j}}
\textcolor{red}{\bcontraction[2ex]{\{p^\dagger q^\dagger sr\}}{a}{^\dagger b^\dagger j}{i}}
\textcolor{red}{\contraction[1ex]{\{p^\dagger q^\dagger}{s}{r\} }{a}}
\textcolor{red}{\contraction[2ex]{\{p^\dagger q^\dagger s}{r}{\} a^\dagger}{b}}
\textcolor{red}{\contraction[3ex]{\{p^\dagger}{q}{^\dagger sr\} a^\dagger b^\dagger}{j}}
\textcolor{red}{\contraction[4ex]{\{}{p}{^\dagger q^\dagger sr\} a^\dagger b^\dagger j}{i}}
\{p^\dagger q^\dagger sr\} a^\dagger b^\dagger ji
+
\textcolor{red}{\bcontraction[1ex]{\{p^\dagger}{q}{^\dagger}{s}}
\textcolor{blue}{\bcontraction[2ex]{\{}{p}{^\dagger q^\dagger s}{r}}
\textcolor{red}{\bcontraction[1ex]{\{p^\dagger q^\dagger sr\} a^\dagger}{b}{^\dagger}{j}}
\textcolor{blue}{\bcontraction[2ex]{\{p^\dagger q^\dagger sr\}}{a}{^\dagger b^\dagger j}{i}}
\textcolor{blue}{\contraction[1ex]{\{p^\dagger q^\dagger s}{r}{\}}{a}}
\textcolor{red}{\contraction[2ex]{\{p^\dagger q^\dagger}{s}{r\} a^\dagger}{b}}
\textcolor{red}{\contraction[3ex]{\{p^\dagger}{q}{^\dagger sr\} a^\dagger b^\dagger}{j}}
\textcolor{blue}{\contraction[4ex]{\{}{p}{^\dagger q^\dagger sr\} a^\dagger b^\dagger j}{i}}
\{p^\dagger q^\dagger sr\} a^\dagger b^\dagger ji \\ 
&= 
\textcolor{blue}{2}\cdot\textcolor{red}{2}\cdot\delta_{pj}\delta_{qi}\delta_{rb}\delta_{sa}
- \textcolor{red}{2}\cdot\delta_{pj}\delta_{qi}\delta_{ra}\delta_{sb} \\ 
&\phantom{=}
- \textcolor{red}{2}\cdot\delta_{pi}\delta_{qj}\delta_{rb}\delta_{sa}
+ \textcolor{blue}{2}\cdot\textcolor{red}{2}\cdot\delta_{pi}\delta_{qj}\delta_{ra}\delta_{sb}\,,
\end{align*} 
where the merged spin paths were highlighted in red and blue. The final terms contributing to $E_{\text{corr}}$ are therefore
\begin{equation*}
E_{\text{corr}} \longleftarrow \sum_{i<j, a<b}\left(4o_{ab}^{ij} - 2o_{ba}^{ij}\right)t_{ij}^{ab}\,.
\end{equation*}

Clearly, due to equations (\ref{wick_1.eq}) and (\ref{wick_2.eq}), there are exceptions for active indices. Every contracted active index either involves only $\alpha$ (for $\contraction{}{v}{^\dagger }{w}v^\dagger w$) or $\beta$ (for $\contraction{}{v}{}{w}vw^\dagger$) spins. Therefore, every path containing at least one active index contributes a prefactor of $1$ instead. The expansion of $\left(\left\{\hat{E}_{rs}^{pq}\right\}\hat{E}_{ij}^{av}\right)_C$, e.g., yields 
\begin{align*}
\left(\left\{\hat{E}_{rs}^{pq}\right\}\hat{E}_{ij}^{av}\right)_C &= 
\bcontraction[1ex]{\{p^\dagger}{q}{^\dagger}{s}
\textcolor{red}{\bcontraction[2ex]{\{}{p}{^\dagger q^\dagger s}{r}}
\textcolor{red}{\bcontraction[1ex]{\{p^\dagger q^\dagger sr\} a^\dagger}{v}{^\dagger}{j}}
\bcontraction[2ex]{\{p^\dagger q^\dagger sr\}}{a}{^\dagger v^\dagger j}{i}
\contraction[1ex]{\{p^\dagger q^\dagger}{s}{r\} }{a}
\textcolor{red}{\contraction[2ex]{\{p^\dagger q^\dagger s}{r}{\} a^\dagger}{v}}
\textcolor{red}{\contraction[3ex]{\{}{p}{^\dagger q^\dagger sr\} a^\dagger v^\dagger}{j}}
\contraction[4ex]{\{p^\dagger}{q}{^\dagger sr\} a^\dagger v^\dagger j}{i}
\{p^\dagger q^\dagger sr\} a^\dagger v^\dagger ji
+
\textcolor{red}{\bcontraction[1ex]{\{p^\dagger}{q}{^\dagger}{s}}
\textcolor{red}{\bcontraction[2ex]{\{}{p}{^\dagger q^\dagger s}{r}}
\textcolor{red}{\bcontraction[1ex]{\{p^\dagger q^\dagger sr\} a^\dagger}{v}{^\dagger}{j}}
\textcolor{red}{\bcontraction[2ex]{\{p^\dagger q^\dagger sr\}}{a}{^\dagger v^\dagger j}{i}}
\textcolor{red}{\contraction[1ex]{\{p^\dagger q^\dagger s}{r}{\}}{a}}
\textcolor{red}{\contraction[2ex]{\{p^\dagger q^\dagger}{s}{r\} a^\dagger}{v}}
\textcolor{red}{\contraction[3ex]{\{}{p}{^\dagger q^\dagger sr\} a^\dagger v^\dagger}{j}}
\textcolor{red}{\contraction[4ex]{\{p^\dagger}{q}{^\dagger sr\} a^\dagger v^\dagger j}{i}}
\{p^\dagger q^\dagger sr\} a^\dagger v^\dagger ji \\ 
&\phantom{=} 
+
\textcolor{red}{\bcontraction[1ex]{\{p^\dagger}{q}{^\dagger}{s}}
\textcolor{red}{\bcontraction[2ex]{\{}{p}{^\dagger q^\dagger s}{r}}
\textcolor{red}{\bcontraction[1ex]{\{p^\dagger q^\dagger sr\} a^\dagger}{v}{^\dagger}{j}}
\textcolor{red}{\bcontraction[2ex]{\{p^\dagger q^\dagger sr\}}{a}{^\dagger v^\dagger j}{i}}
\textcolor{red}{\contraction[1ex]{\{p^\dagger q^\dagger}{s}{r\} }{a}}
\textcolor{red}{\contraction[2ex]{\{p^\dagger q^\dagger s}{r}{\} a^\dagger}{v}}
\textcolor{red}{\contraction[3ex]{\{p^\dagger}{q}{^\dagger sr\} a^\dagger v^\dagger}{j}}
\textcolor{red}{\contraction[4ex]{\{}{p}{^\dagger q^\dagger sr\} a^\dagger v^\dagger j}{i}}
\{p^\dagger q^\dagger sr\} a^\dagger v^\dagger ji
+
\textcolor{red}{\bcontraction[1ex]{\{p^\dagger}{q}{^\dagger}{s}}
\bcontraction[2ex]{\{}{p}{^\dagger q^\dagger s}{r}
\textcolor{red}{\bcontraction[1ex]{\{p^\dagger q^\dagger sr\} a^\dagger}{v}{^\dagger}{j}}
\bcontraction[2ex]{\{p^\dagger q^\dagger sr\}}{a}{^\dagger v^\dagger j}{i}
\contraction[1ex]{\{p^\dagger q^\dagger s}{r}{\}}{a}
\textcolor{red}{\contraction[2ex]{\{p^\dagger q^\dagger}{s}{r\} a^\dagger}{v}}
\textcolor{red}{\contraction[3ex]{\{p^\dagger}{q}{^\dagger sr\} a^\dagger v^\dagger}{j}}
\contraction[4ex]{\{}{p}{^\dagger q^\dagger sr\} a^\dagger v^\dagger j}{i}
\{p^\dagger q^\dagger sr\} a^\dagger v^\dagger ji \\ 
&= 2\cdot \textcolor{red}{1} \cdot\delta_{pj}\delta_{qi}\delta_{rv}\delta_{sa} - \textcolor{red}{1}\cdot\delta_{pj}\delta_{qi}\delta_{ra}\delta_{sv} \\ 
&\phantom{=} - \textcolor{red}{1} \cdot\delta_{pi}\delta_{qj}\delta_{rv}\delta_{sa} + 2\cdot\textcolor{red}{1}\cdot\delta_{pi}\delta_{qj}\delta_{ra}\delta_{sv}\,.
\end{align*} 
where merged spin paths of active indices are highlighted in red.

Lastly, consider the example:
\begin{align*}
\left(\left\{\hat{E}_{rs}^{pq}\right\}\hat{E}_{iw}^{va}\right)_C &= 
\bcontraction[1ex]{\{p^\dagger}{q}{^\dagger}{s}
\bcontraction[2ex]{\{}{p}{^\dagger q^\dagger s}{r}
\bcontraction[1ex]{\{p^\dagger q^\dagger sr\} v^\dagger}{a}{^\dagger}{w}
\bcontraction[2ex]{\{p^\dagger q^\dagger sr\}}{v}{^\dagger a^\dagger w}{i}
\contraction[1ex]{\{p^\dagger q^\dagger}{s}{r\} }{v}
\contraction[2ex]{\{p^\dagger q^\dagger s}{r}{\} v^\dagger}{a}
\contraction[3ex]{\{}{p}{^\dagger q^\dagger sr\} v^\dagger a^\dagger}{w}
\contraction[4ex]{\{p^\dagger}{q}{^\dagger sr\} v^\dagger a^\dagger w}{i}
\{p^\dagger q^\dagger sr\} v^\dagger a^\dagger wi
+
\cancel{\bcontraction[1ex]{\{p^\dagger}{q}{^\dagger}{s}
\bcontraction[2ex]{\{}{p}{^\dagger q^\dagger s}{r}
\bcontraction[1ex]{\{p^\dagger q^\dagger sr\} v^\dagger}{a}{^\dagger}{w}
\bcontraction[2ex]{\{p^\dagger q^\dagger sr\}}{v}{^\dagger a^\dagger w}{i}
\contraction[1ex]{\{p^\dagger q^\dagger s}{r}{\}}{v}
\contraction[2ex]{\{p^\dagger q^\dagger}{s}{r\} v^\dagger}{a}
\contraction[3ex]{\{}{p}{^\dagger q^\dagger sr\} v^\dagger a^\dagger}{w}
\contraction[4ex]{\{p^\dagger}{q}{^\dagger sr\} v^\dagger a^\dagger w}{i}
\{p^\dagger q^\dagger sr\} v^\dagger a^\dagger wi} \\ 
&\phantom{=} 
+
\cancel{\bcontraction[1ex]{\{p^\dagger}{q}{^\dagger}{s}
\bcontraction[2ex]{\{}{p}{^\dagger q^\dagger s}{r}
\bcontraction[1ex]{\{p^\dagger q^\dagger sr\} v^\dagger}{a}{^\dagger}{w}
\bcontraction[2ex]{\{p^\dagger q^\dagger sr\}}{v}{^\dagger a^\dagger w}{i}
\contraction[1ex]{\{p^\dagger q^\dagger}{s}{r\} }{v}
\contraction[2ex]{\{p^\dagger q^\dagger s}{r}{\} v^\dagger}{a}
\contraction[3ex]{\{p^\dagger}{q}{^\dagger sr\} v^\dagger a^\dagger}{w}
\contraction[4ex]{\{}{p}{^\dagger q^\dagger sr\} v^\dagger a^\dagger w}{i}
\{p^\dagger q^\dagger sr\} v^\dagger a^\dagger wi}
+
\bcontraction[1ex]{\{p^\dagger}{q}{^\dagger}{s}
\bcontraction[2ex]{\{}{p}{^\dagger q^\dagger s}{r}
\bcontraction[1ex]{\{p^\dagger q^\dagger sr\} v^\dagger}{a}{^\dagger}{w}
\bcontraction[2ex]{\{p^\dagger q^\dagger sr\}}{v}{^\dagger a^\dagger w}{i}
\contraction[1ex]{\{p^\dagger q^\dagger s}{r}{\}}{v}
\contraction[2ex]{\{p^\dagger q^\dagger}{s}{r\} v^\dagger}{a}
\contraction[3ex]{\{p^\dagger}{q}{^\dagger sr\} v^\dagger a^\dagger}{w}
\contraction[4ex]{\{}{p}{^\dagger q^\dagger sr\} v^\dagger a^\dagger w}{i}
\{p^\dagger q^\dagger sr\} v^\dagger a^\dagger wi \\ 
&= \delta_{pw}\delta_{qi}\delta_{sv}\delta_{ra} + \delta_{pi}\delta_{qw}\delta_{rv}\delta_{sa}
\end{align*}
Whenever the two active indices $v$ and $w$ are contracted as spin orbital particles (e.g. $\contraction{}{r}{}{v}rv^\dagger$) and holes (e.g. $\contraction{}{p}{^\dagger }{w}p^\dagger w$) on the same spin path, the whole expression vanishes. This is due to incompatible active spins (c.f. equation \ref{wick_2.eq}). While the creator -- annihilator contraction requires $\alpha$ spins, the annihilator -- creator contraction requires $\beta$ spins. Both can never be true at the same time. An analogue argument holds for the second one-particle interrelating operator $\hat{F}_N^{(2)}$ containing spin orbital $\alpha\longrightarrow\alpha$ substitutions only. Here, any active creator -- annihilator contractions on the same spin path lead to a vanishing expression. 

In this work, the BCH expansion of the CCSD equations (\ref{CCEquations.eq}) up to quadruply nested commutators was generated. Please note that due to the non-vanishing $\contraction{}{\hat{T}}{}{\hat{T}}\hat{T}\hat{T}$ contractions, the BCH series does not naturally truncate at fourth order. We restrict our CCSD approach to BCH order four, nevertheless, since fifth, sixth, etc. order terms are expected to have negligible contributions, i.e. $<10^{-7}\,$a.u. in terms of the correlation energy. We will further discuss and motivate this issue in section \ref{Application}.

\subsection{Representation by spin-free Goldstone diagrams}
\label{Goldstone}

Application of Wick's theorem to the BCH expansion of the CCSD equations (\ref{CCEquations.eq}) leads to largely redundant products of Kronecker deltas. These bind general indices of $\hat{H}_N$ to specific indices from $\hat{T}$ or the external projections. 

Due to the $\hat{T}$ and $\hat{H}_N$ tensor symmetries, many Wick terms lead to redundant equations. A much more efficient approach to CC equation derivation -- established by \v{C}\'{i}\v{z}ek and Paldus\cite{Cizek1966,Paldus1975,Paldus1977} -- is based on the diagrammatic representations developed by Goldstone\cite{Goldstone1957} and Hugenholtz\cite{Hugenholtz1957}, who generalized Feynman's particle interaction diagrams\cite{Feynman1949} to many-body perturbation theory. 

In this work, we apply special spin-free Goldstone diagrams to identify summarizable terms. The general idea is to abstract algebraically redundant terms into isomorphic graphs such that the redundancy check may be performed by the optimized (sub)graph isomorphism testing algorithm VF2\cite{Cordella2001,Cordella2004} contained in the boost graph library (BGL)\cite{BGLBook}. In the following, all essential details of the used diagrams are given. Firstly, the new lines and features to properly represent active and/or identical lines are introduced (c.f. subsection \ref{newfeatures}). Then, in subsection \ref{tensorsymmetry}, the embedding of tensor symmetries without implicit antisymmetrization is motivated and finally, in subsection \ref{indexrelations}, the incorporation of index relations into the diagram's topology is discussed.

\subsubsection{New diagram features / diagram augmentation}
\label{newfeatures}

To properly represent active indices in Goldstone diagrams, a new type of line needs to be introduced. Since active index operators may contract as either holes or particles (c.f. equation \ref{wick_2.eq}), the vertical orientation of their diagrammatic representation not only needs to be arbitrary but also needs to be changeable. To accommodate this, we represent active indices by particle/hole lines with hollow arrows. The hollowness shall emphasize the lines not being \glqq{p}inned\grqq{} to the surrounding background such that they may change their vertical direction freely. For the operator $\hat{E}_{iw}^{va}$, e.g., the Goldstone diagram representation is given by 
\begin{center}
\begin{tabular}{ccc} 
$\hat{E}_{iw}^{va}$ & $=$ & \cincludegraphics[scale=0.75]{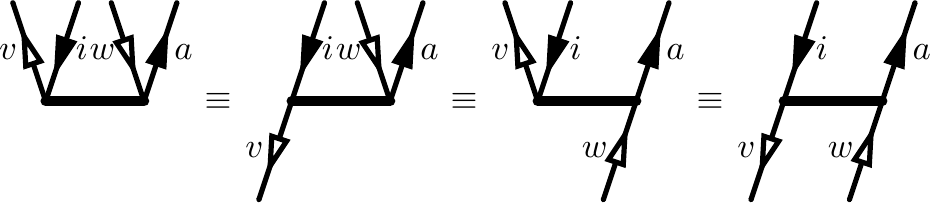}\,,
\end{tabular}
\end{center}
where indices $v$ and $w$ are represented by hollow-arrowed lines with arbitrary/fluid vertical direction. 

For spectating (e.g. $\hat{E}_{vi}^{av}$) or identical (e.g. $\hat{E}_{ii}^{aa}$) indices, the pairwise line affiliation needs to be part of the diagram. These pairwise identical indices will be represented by background color-coded lines, where same color indicates same indices: 
\begin{center}
\begin{tabular}{ccccccc} 
$\hat{E}_{iv}^{va}$ & $=$ & \cincludegraphics[scale=0.75]{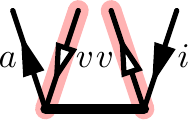} & \hspace{1cm},\hspace{1cm} & $\hat{E}_{ii}^{aa}$ & $=$ & \cincludegraphics[scale=0.75]{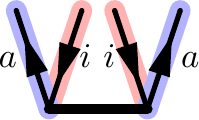}
\end{tabular}
\end{center}

It is obvious that the contraction rules for active indices (c.f. equation \ref{wick_2.eq}) are automatically fulfilled by the connection of heads and tails of hollow-arrowed lines analogous to the connection of particle/hole lines. However, some non-trivial findings occur when building CC diagrams in the presented SASC-CC framework:

\begin{itemize}
\item[(i)] due to the operator generation scheme\cite{Herrmann2020a}, which handles spectating $\hat{E}$ operators explicitly, all non-vanishing $\contraction{}{\phantom{\rule{2pt}{8pt}}}{}{\hspace{8pt}}\hat{T}$ contractions can occur among identical active indices, i.e. spectators, leading to active snakes, only. All other non-spectating operators possess either $<$ or $\neq$ relations between their active indices such that no accidental spectators are formable. Consider, e.g., the following term  contributing to $\braket{\Psi_0|\hat{F}_N^{(1)}\hat{T}_1\Psi_0}$:
\begin{tabular}{ccc}
$\left(\hat{F}_N^{(1)}\hat{E}_{vi}^{av}\right)_{FC}$ & \hspace{5pt}$=$\hspace{5pt} & \cincludegraphics[scale=0.75]{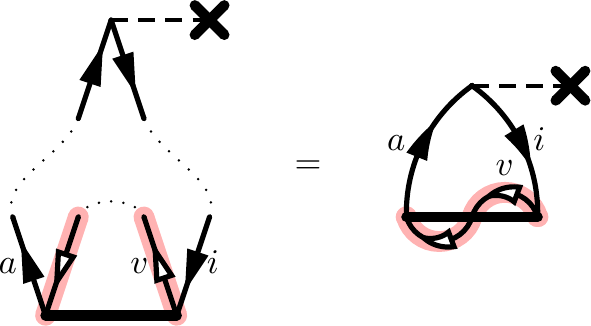}
\end{tabular}
Here, the two active lines corresponding to the spectating index $v$ form an internally connected snake, which connects the two vertices of the cluster line. The final contribution to $E_{\text{corr}}$ therefore yields
\begin{equation*}
\braket{\Psi_0|\hat{F}_N^{(1)}\hat{T}_1\Psi_0} \longleftarrow -\sum_{iav}f_a^i t_{vi}^{av}\,.
\end{equation*}
At the same time, no contribution to $E_{\text{corr}}$ is obtained from $\left(\hat{F}_N^{(1)}\hat{E}_{iw}^{va}\right)_{FC}$ because indices $v$ and $w$ may never be identical in $\sum_{ia}\sum_{v\neq w}\hat{E}_{iw}^{va}$ (c.f. equation \ref{Doubles.eq}):
\begin{tabular}{ccc}
$\left(\hat{F}_N^{(1)}\hat{E}_{iw}^{va}\right)_{FC}$ & \hspace{5pt}$=$\hspace{5pt} & \cincludegraphics[scale=0.75]{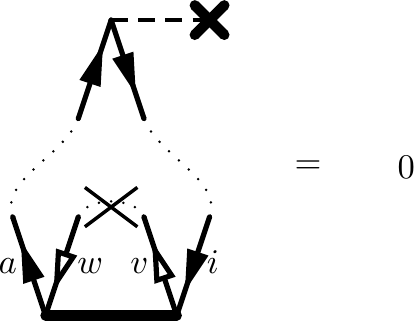}
\end{tabular}

\item[(ii)] when comparing topologies of spin-free Goldstone diagrams, the direction of the participating active lines must not be taken into account. This is easily motivated by a small example. Consider the terms $A$ and $B$ with 
\begin{align*}
A &= \frac{1}{2}\sum_{iavw}t_i^vt_w^a\braket{\Psi_0|\hat{F}_N^{(1)}\hat{E}_i^v\hat{E}_w^a\Psi_0} \\
&= \frac{1}{2}\sum_{pq}\sum_{iavw}f_q^pt_i^vt_w^a 
\bcontraction{\{}{p}{^\dagger }{q} 
\bcontraction{\{p^\dagger q\}}{v}{^\dagger }{i} 
\bcontraction{\{p^\dagger q\}v^\dagger i}{a}{^\dagger }{w}
\contraction[1ex]{\{p^\dagger q\}}{v}{^\dagger ia^\dagger }{w}
\contraction[2ex]{\{p^\dagger }{q}{\}v^\dagger i}{a}
\contraction[3ex]{\{}{p}{^\dagger q\}v^\dagger }{i}
\{p^\dagger q\}v^\dagger ia^\dagger w \\
&= -\frac{1}{2}\sum_{iav}f_a^it_i^vt_v^a
\end{align*}
and
\begin{align*}
B &= \frac{1}{2}\sum_{iavw}t_w^at_i^v\braket{\Psi_0|\hat{F}_N^{(1)}\hat{E}_w^a\hat{E}_i^v\Psi_0} \\
&= \frac{1}{2}\sum_{pq}\sum_{iavw}f_q^pt_w^at_i^v 
\bcontraction{\{p^\dagger q\}a^\dagger w}{v}{^\dagger }{i} 
\bcontraction{\{p^\dagger q\}}{a}{^\dagger }{w} 
\bcontraction{\{}{p}{^\dagger }{q}
\contraction[1ex]{\{p^\dagger q\}a^\dagger }{w}{}{v}
\contraction[1ex]{\{p^\dagger }{q}{\}}{a}
\contraction[2ex]{\{}{p}{^\dagger q\}a^\dagger wv^\dagger }{i}
\{p^\dagger q\}a^\dagger wv^\dagger i \\
&= +\frac{1}{2}\sum_{iav}f_a^it_v^at_i^v\,,
\end{align*}
where indices $v$ and $w$ connect as spin orbital holes ($A$: creator -- annihilator) and as spin orbital particles ($B$: annihilator -- creator), respectively. Clearly, in the full CCSD energy equation it is $E_{\text{corr}} \longleftarrow A+B = 0$ due to 
\begin{equation*}
\contraction{\left[\right.}{\hat{E}}{_i^v, }{\hat{E}}
\left[\hat{E}_i^v, \hat{E}_w^a \right]_{+} = 0\,,
\end{equation*}
where $\left[\ldots\right]_+$ shall denote the anticommutator.
In order to find these kinds of redundancies graphically, i.e. identify the Goldstone diagrams of A and B as topologically equivalent, all information about the vertical direction of the active line has to be neglected:\\
\begin{tabular}{ccc}
\phantom{blabla} & \cincludegraphics[scale=0.75]{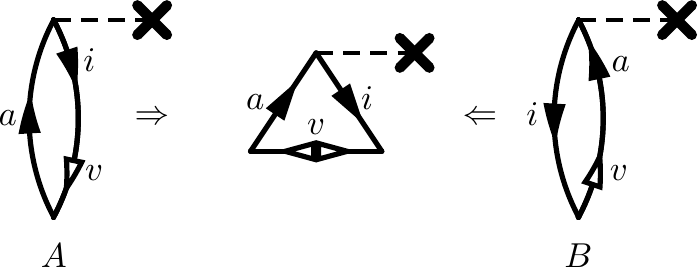} &\phantom{blabla}
\end{tabular} \vspace{5pt} \\ 
While this certainly leads to isomorphic Goldstone diagrams $A$ and $B$, it also removes information about the sign of the diagrams. In our case, however, this is of no concern because the sign (and also the prefactor) is already apparent from Wick's theorem. As stated before, we merely apply Goldstone diagram representations to identify redundant terms.   

\item[(iii)] Whenever identical indices are contracted to different externals, the whole expression vanishes. This again results directly from operator generation scheme\cite{Herrmann2020a} applied in this work, where index relations were explicitly derived (also for external indices). As an example, consider the doubles projection $\left(\hat{E}_{AB}^{IJ}\hat{E}_{ii}^{aa}\right)_{FC}$ with
\begin{equation*}
\sum_{ia}\braket{\hat{E}_{IJ}^{AB}\Psi_0|\hat{E}_{ii}^{aa}\Psi_0} = 0 \qquad\forall_{I<J,\,A<B}\,.
\end{equation*}
Due to the contractions $\contraction{}{I}{^\dagger }{i}I^\dagger i = \delta_{Ii}$ and $\contraction{}{J}{^\dagger }{i}J^\dagger i = \delta_{Ji}$ never being compatible with $I<J$, the term vanishes. Please note that the same argument also holds for the particle contractions $\contraction{}{A}{}{a}Aa^\dagger = \delta_{Aa}$ and $\contraction{}{B}{}{a}Ba^\dagger = \delta_{Ba}$ in this case. Further remarks on the incorporation of index relations are given in subsection \ref{indexrelations}.

\item[(iv)] Visually, the second one-particle interrelating operator $\hat{F}_N^{(2)}$ resembles an internally contracted $\hat{V}_N$ interaction line: \vspace{5pt} \\ 
\begin{tabular}{ccc}
$\hat{F}_{N}^{(2)}$ & $=$ & \cincludegraphics[scale=0.72]{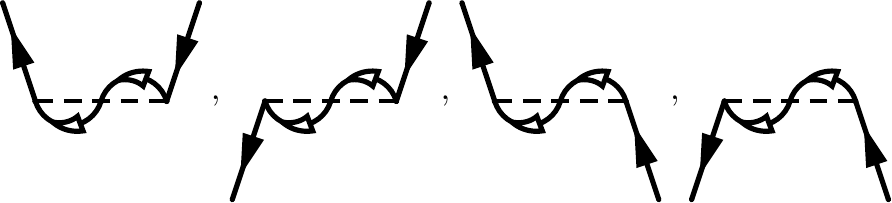}
\end{tabular} \vspace{5pt} \\ 
As mentioned in (ii) the sign and prefactor of the diagrams are evaluated from Wick's theorem. A diagrammatic representation of the $\alpha \longrightarrow \alpha$ only substitution of $\hat{F}_N^{(2)}$ is therefore not required. 

\end{itemize}

\subsubsection{Incorporating tensor symmetry}
\label{tensorsymmetry}

In contrast to spin orbital substitution operators $\hat{X}$, the spatial substitutions $\hat{E}$ possess a pairwise permutational symmetry only. That is 
\begin{equation}
\hat{E}_{p_1p_2\ldots p_\nu}^{q_1q_2\ldots q_\nu} = \hat{E}_{p_2p_1\ldots p_\nu}^{q_2q_1\ldots q_\nu} = \ldots = \hat{E}_{\hat{P}\left(p_1p_2\ldots p_\nu\right)}^{\hat{P}\left(q_1q_2\ldots q_\nu\right)} \quad \forall_{\hat{P} \in \mathbb{S}_\nu}\,,
\end{equation}

where $\hat{P}$ shall denote an arbitrary permutation from the symmetric group $\mathbb{S}_\nu$. A lot of redundancy in the generated Wick terms, originates from this symmetry. Unfortunately, Goldstone diagrams can not completely incorporate the latter. The Goldstone representation of $\hat{T}$ operators is a set of horizontally arranged vertices connected by a bold line. Each vertex is associated with an outgoing and an incoming line representing the annihilator/creator pairs of the operator. In Table \ref{Tensor2Goldstone.tab}, tensor and Goldstone representations of all permutational equivalent rank\footnote{Beware that rank implies tensor rank and not substitution order. A rank two operator may very well resemble a spatial substitution of order one, as e.g. in $\hat{E}_{vi}^{av}$.} one to three cluster operators are shown. 

\begin{table}[h]
\caption{Permutational symmetry of tensor and Goldstone representations of cluster operators with tensor ranks one to three. The cluster vertices are colored to increase readability.}
\label{Tensor2Goldstone.tab} 
\begin{tabular}{c|c}
\textbf{Tensors} & \textbf{Goldstone diagrams} \\ \hline
\rule{0pt}{20pt}$\hat{E}_{p_1}^{q_1}$ & \cincludegraphics[scale=0.75]{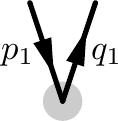} \\[10pt] \hline 
\rule{0pt}{25pt}$\hat{E}_{p_1p_2}^{q_1q_2} = \hat{E}_{p_2p_1}^{q_2q_1}$ & \cincludegraphics[scale=0.75]{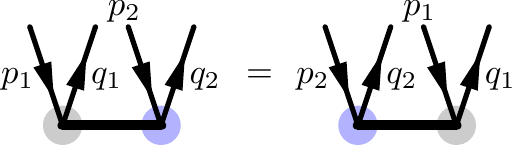} \\[15pt] \hline 
\rule{0pt}{25pt}$\hat{E}_{p_1p_2p_3}^{q_1q_2q_3} = \hat{E}_{p_3p_2p_1}^{q_3q_2q_1}$ & \cincludegraphics[scale=0.75]{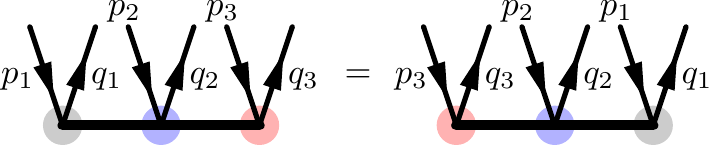} \\[20pt]
\rotatebox{90}{$=$} & \rotatebox{90}{$\neq$} \\
\rule{0pt}{25pt}$\hat{E}_{p_2p_1p_3}^{q_2q_1q_3} = \hat{E}_{p_3p_1p_2}^{q_3q_1q_2}$ & \cincludegraphics[scale=0.75]{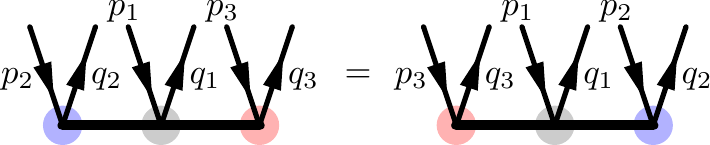} \\[20pt]
\rotatebox{90}{$=$} & \rotatebox{90}{$\neq$} \\
\rule{0pt}{25pt}$\hat{E}_{p_1p_3p_2}^{q_1q_3q_2} = \hat{E}_{p_2p_3p_1}^{q_2q_3q_1}$ & \cincludegraphics[scale=0.75]{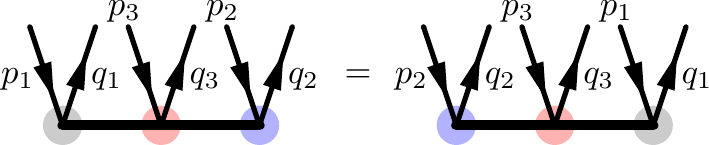} 
\end{tabular}
\end{table}

The only symmetric feature of the shown non-antisymmetrized Goldstone diagrams is the spatial inversion with respect to the center of the cluster line. As clearly visible in Table \ref{Tensor2Goldstone.tab}, for tensor ranks one and two this exactly preserves the right permutational symmetry by accident. While rank one operators do not possess any permutational symmetry, rank two operators possess just one resembling the transposition of indices 1 and 2 (highlighted by blue and gray vertices). This transposition is identical to an inversion such that the two possible diagrams are topologically equivalent, i.e. isomorphic.  

Rank three operators, on the other hand, possess $3! = 6$ equivalent permutations. By graphical inversion, only three groups of two diagrams are found isomorphic. To resolve this problem in spin orbital theory, one may neglect all particle information by transitioning from Goldstone to Hugenholtz diagrams, as e.g. in 
\begin{center}
\includegraphics[scale=0.75]{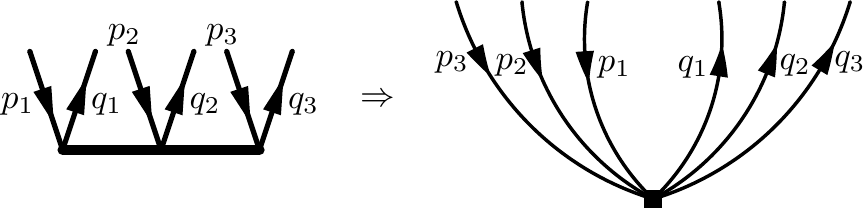}
\end{center}
where all cluster vertices (of the same cluster line) are merged into a single node. While this transition clearly allows for the required permutational tensor symmetry, it also destroys any information about the pairwise creator/annihilator affiliation. In the spin orbital case, this is of no concern since all participating tensors may be antisymmetrized. In the SASC approach followed in this work, however, it is particularly important to use non-antisymmetrized Goldstone diagrams (as e.g. also discussed in\cite{Datta2013}). Therefore, the transition to Hugenholtz diagrams is not a viable option.

A different approach to achieve the correct permutational symmetry, is to replace cluster lines by a totally connected, i.e. complete, subgraph. In general, a complete graph $K_n$ is defined by a unique set of $n$ vertices, where each unique pair of vertices is connected by a unique undirected edge. Such graphs are therefore topologically invariant to arbitrary vertex permutations. 

In this work, we represent all occuring substitution operators by such totally connected cluster lines. 
When representing the latter graphically by means of Goldstone diagrams, however, a lot of complexity is introduced through the additional cluster lines. To remove this unnecessary complexity, we abbreviate fully connected cluster lines by ring-shaped cluster lines we denote \textit{cluster super lines}. In Table \ref{Complete2Superline.tab}, fully connected Goldstone representations as well as their cluster super line abbreviation of rank one to four substitution operators are shown.

\begin{table}[h]
\caption{\label{Complete2Superline.tab}Fully connected cluster lines and their cluster super line representation for rank one to four substitution operators.}
\begin{tabular}{c|ccc} 
\textbf{Rank} & \textbf{Fully connected} & & \textbf{Super line} \\ \hline

\rule{0pt}{19pt}1 & \cincludegraphics[scale=0.75]{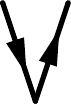} & $\equiv$ & \cincludegraphics[scale=0.75]{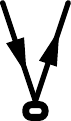}\\[10pt] \hline
\rule{0pt}{19pt}2 & \cincludegraphics[scale=0.75]{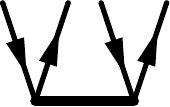} & $\equiv$ & \cincludegraphics[scale=0.75]{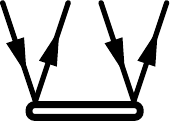}\\[10pt] \hline
\rule{0pt}{23pt}3 & \cincludegraphics[scale=0.75]{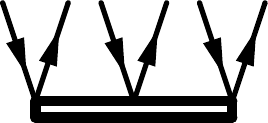} & $\equiv$ & \cincludegraphics[scale=0.75]{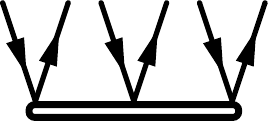}\\[10pt] \hline
\rule{0pt}{23pt}4 & \cincludegraphics[scale=0.75]{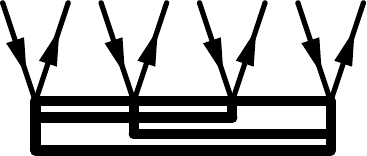} & $\equiv$ & \cincludegraphics[scale=0.75]{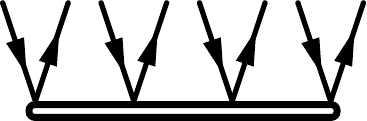}
\end{tabular}
\end{table}

\subsubsection{Incorporating index relations}
\label{indexrelations}

Spin-complete and linear-independent substitution operators were generated by the use of L\"owdin's projection operator method\cite{Herrmann2020a}. One of the major consequences of this generative approach is the presence of index relations (orderings) in all substitution operators. By definition\cite{Herrmann2020a}, it is 
\begin{itemize}
\item $i < j < k < \ldots \quad \in \mathbb{O}$
\item $a < b < c < \ldots \quad \in \mathbb{A}$
\item active creators: $v^{(i)} < w^{(i)} < \ldots \quad \in \mathbb{A}^{(i)}$
\item active annihilators: $v^{(ii)} < w^{(ii)} < \ldots \quad \in \mathbb{A}^{(ii)}$
\item active spectators: $v^{(iii)} < w^{(iii)} < \ldots \quad \in \mathbb{A}^{(iii)}$
\item $\mathbb{A}^{(i)} \cap \mathbb{A}^{(ii)} = \mathbb{A}^{(i)} \cap \mathbb{A}^{(iii)} = \mathbb{A}^{(ii)} \cap \mathbb{A}^{(iii)} = \emptyset\,,$
\end{itemize}
for the index set of each individual substitution operator. 

While this is beneficial in guaranteeing the exactly required amount of amplitudes, it enforces us to distinguish between e.g. $t_{ij}^{ab}$ and $t_{ji}^{ab}$ in  
\begin{equation*}
\hat{T}_2 = \ldots \sum_{i<j, a<b} t_{ij}^{ab}\hat{E}_{ij}^{ab} + \sum_{i<j, a<b} t_{ji}^{ab}\hat{E}_{ji}^{ab} + \ldots\,.
\end{equation*}

Clearly, however, the unlabeled Goldstone diagrams resulting from these two operators are identical: 
\begin{center}
\begin{tabular}{ccccc}
$\hat{E}_{ij}^{ab}$ & $=$ & \cincludegraphics[scale=0.75]{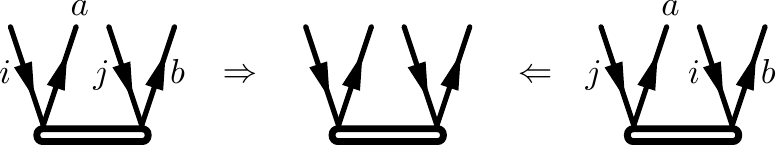} & $=$ & $\hat{E}_{ji}^{ab}$
\end{tabular}
\end{center} 

To resolve this problem, we employed a simple solution. For each particle/active/hole line connected to a cluster super line, we place auxiliary vertices in between the cluster vertices and the contracted lines. The labels of these vertices shall then reflect the index relations within the particle/active/hole domain of the associated cluster super line (via 1 < 2 < 3 etc.). Since every cluster super line possesses its own set of auxiliary vertices, their labels do not need to be updated while contracting. This ensures a fast diagram generation, which is particularly important for the large number of occurring Wick terms. At this stage we only aim to find redundant Wick terms and do not consider, e.g., vanishing terms due to impossible index symmetries. In this fashion, the evaluation of the final (merged) index relations is only performed for the collected (non-redundant) diagrams.

For the given examples of $t_{ij}^{ab}$ and $t_{ji}^{ab}$, consider the terms
\begin{center}
\begin{tabular}{ccccc}
$\sum\limits_{i<j,a<b}\braket{\Psi_0|\hat{V}_Nt_{ij}^{ab}\hat{E}_{ij}^{ab}\Psi_0}$ & $\Rightarrow$ & \cincludegraphics[scale=0.75]{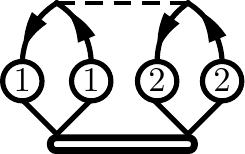} & $+$ & \cincludegraphics[scale=0.75]{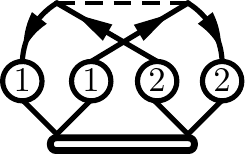} \\[15pt]
& $=$ & \multicolumn{3}{c}{$\sum\limits_{i<j,a<b}\left(4o_{ab}^{ij} - 2o_{ba}^{ij}\right)t_{ij}^{ab}$} \\[15pt]
$\sum\limits_{i<j,a<b}\braket{\Psi_0|\hat{V}_Nt_{ji}^{ab}\hat{E}_{ji}^{ab}\Psi_0}$ & $\Rightarrow$ & \cincludegraphics[scale=0.75]{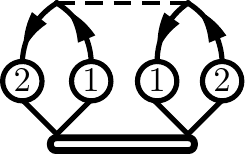} & $+$ & \cincludegraphics[scale=0.75]{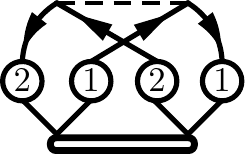} \\[15pt]
& $=$ & \multicolumn{3}{c}{$\sum\limits_{i<j,a<b}\left(-2o_{ab}^{ij} + 4o_{ba}^{ij}\right)t_{ji}^{ab}$}
\end{tabular}
\end{center}
where topologically equivalent diagrams are only shown once. Due to the newly added auxiliary vertices, the four diagrams are now trivially distinct. 

As a final example, consider the term 
\begin{equation*}
\frac{1}{4}\sum_{pqrs}\sum_{\substack{i<j \\ k}}\sum_{\substack{a \\ b<c}}\sum_{\substack{v\neq w \\ x}} o_{rs}^{pq}t_{wji}^{avw}t_{kx}^{bc}\braket{\hat{E}_{VI}^{AV}\Psi_0| \hat{E}_{rs}^{pq}\hat{E}_{wji}^{avw}\hat{E}_{kx}^{bc}\Psi_0}
\end{equation*}
contributing to the CCSD singles equation. One of the possible diagrams for this term is given by 

\onecolumngrid
\begin{center}
\begin{table}[h]
\caption{\label{E2T2.tab}Progression of the index relation graphs of the contraction pattern given in equation (\ref{E2T2.eq}). The merged indices of each step are colored.}
\begin{tabular}{c|c|c|c}
\textbf{Index space} &
$\bcontraction[1ex]{I^\dagger }{J}{^\dagger}{B}
\bcontraction[2ex]{}{I}{^\dagger J^\dagger B}{A}
I^\dagger J^\dagger BA
\quad\quad\quad\quad
\bcontraction[1ex]{a^\dagger }{b}{^\dagger }{j}
\bcontraction[2ex]{}{a}{^\dagger b^\dagger j}{i}
a^\dagger b^\dagger ji$ 
& 
$\contraction[1ex]{\highlight[myred]{I}^\dagger J^\dagger B}{\highlight[myblue]{A}}{\quad\quad\quad\quad a^\dagger }{\highlight[myblue]{b}}
\contraction[4ex]{}{\highlight[myred]{I}}{^\dagger J^\dagger B\highlight[myblue]{A}\quad\quad\quad\quad a^\dagger \highlight[myblue]{b}^\dagger j}{\highlight[myred]{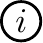}}
\bcontraction[1ex]{\highlight[myred]{I}^\dagger }{J}{^\dagger}{B}
\bcontraction[2ex]{}{\highlight[myred]{I}}{^\dagger J^\dagger B}{\highlight[myblue]{A}}
\highlight[myred]{I}^\dagger J^\dagger B\highlight[myblue]{A}
\quad\quad\quad\quad
\bcontraction[1ex]{a^\dagger }{\highlight[myblue]{b}}{^\dagger }{j}
\bcontraction[2ex]{}{a}{^\dagger \highlight[myblue]{b}^\dagger j}{\highlight[myred]{i}}
a^\dagger \highlight[myblue]{b}^\dagger j\highlight[myred]{i}$ 
&
$\contraction[1.5ex]{I^\dagger \highlight[mygreen]{J}^\dagger \highlight[mygray]{B}}{A}{\quad\quad\quad\quad\highlight[mygray]{a}^\dagger }{b}
\contraction[2ex]{I^\dagger \highlight[mygreen]{J}^\dagger }{\highlight[mygray]{B}}{A\quad\quad\quad\quad}{\highlight[mygray]{a}}
\contraction[3ex]{I^\dagger }{\highlight[mygreen]{J}}{^\dagger \highlight[mygray]{B}A\quad\quad\quad\quad\highlight[mygray]{a}^\dagger b^\dagger }{\highlight[mygreen]{j}}
\contraction[4.5ex]{}{I}{^\dagger \highlight[mygreen]{J}^\dagger \highlight[mygray]{B}A\quad\quad\quad\quad\highlight[mygray]{a}^\dagger b^\dagger \highlight[mygreen]{j}}{i}
\bcontraction[1ex]{I^\dagger }{\highlight[mygreen]{J}}{^\dagger}{\highlight[mygray]{B}}
\bcontraction[2ex]{}{I}{^\dagger \highlight[mygreen]{J}^\dagger \highlight[mygray]{B}}{A}
I^\dagger \highlight[mygreen]{J}^\dagger \highlight[mygray]{B}A
\quad\quad\quad\quad
\bcontraction[1ex]{\highlight[mygray]{a}^\dagger }{b}{^\dagger }{\highlight[mygreen]{j}}
\bcontraction[2ex]{}{\highlight[mygray]{a}}{^\dagger b^\dagger \highlight[mygreen]{j}}{i}
\highlight[mygray]{a}^\dagger b^\dagger \highlight[mygreen]{j}i$  \\[10pt] \hline
$\mathbb{O}$\rule{0pt}{25pt} & \cincludegraphics[scale=1]{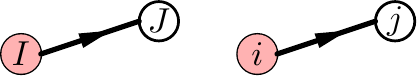} & \cincludegraphics[scale=1]{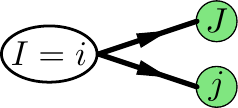} & \cincludegraphics[scale=1]{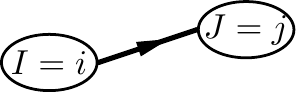} \\[20pt] 
$\mathbb{V}$\rule{0pt}{20pt} & \cincludegraphics[scale=1]{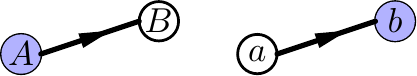} & \cincludegraphics[scale=1]{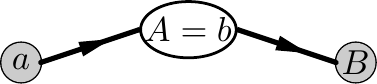} & \cincludegraphics[scale=1]{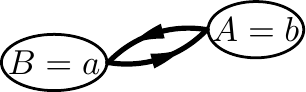} \\[20pt]
$\mathbb{A}$\rule{0pt}{15pt} & -- & -- & --
\end{tabular}
\end{table}
\end{center}
\twocolumngrid

\begin{center}
\includegraphics[scale=0.75]{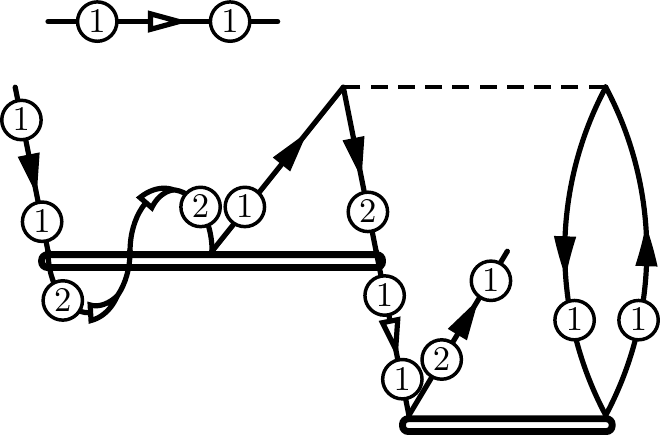}
\end{center}
where all auxiliary vertices are explicitly shown and the external spectator $V$ is contracted to itself. To evaluate the final expression, all index relations need to be merged according to all contracted indices. This evaluation is a non-trivial task, where special care needs to be taken. We will discuss this issue in greater detail throughout the next subsection (\ref{IndexRelations}). For this particular example (collecting all redundant diagrams), the final contribution to the singles residual $R_{VI}^{AV}$ reads 
\begin{equation*}
R_{VI}^{AV} \longleftarrow +1\sum_{\substack{I<i \\ j}}\sum_{\substack{a \\ b < A}}\sum_{\substack{v \neq w}} o_{ab}^{ij} t_{wiI}^{avw}t_{jv}^{bA} \quad \forall_{V\in\mathbb{A}}\,.
\end{equation*}

\subsection{Evaluation of tensor index relations}
\label{IndexRelations}

In the specially crafted Goldstone diagrams presented in subsection \ref{Goldstone}, each pair of contracted auxiliary vertices corresponds to a Kronecker delta merging two distinct index relations. By this union, index relations of one cluster (or external) super line may be coupled to the index relations of another. In this process, impossible relations, such as $i < j < i$, may be constructed thereby producing vanishing diagrams. In this work, we evaluate and keep track of the index relations by the use of a given set of graphs we denote \textit{index relation graphs} (IRGs). 

Each IRG consists of labelled nodes representing spatial indices and directed edges representing a $<$ connection. Clearly, any loop in an IRG -- indicating the presence of both $<$ and $>$ relations at the same time -- constitute an impossible situation. Therefore, valid IRGs are directed acyclic graphs in general. 

The relation $i<j<k, l$ e.g. corresponds to the IRG 
\begin{center}
\includegraphics[scale=1]{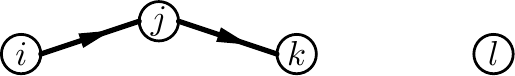}\,,
\end{center}
where index $l$ is not bound to $i,j,k$ at all and therefore represented as an isolated (disconnected) node. 

In the three separated index spaces $\mathbb{O}$, $\mathbb{A}$ and $\mathbb{V}$ employed in this work, only contractions within the same space survive. Therefore, three separate IRGs can be used to represent the index relations of all occupied, active and virtual indices, respectively.

IRGs are particularly useful to evaluate merged index relations for a given contraction pattern. Consider, e.g., the contraction pattern 
\begin{equation}
\label{E2T2.eq}
\braket{\hat{E}_{IJ}^{AB}\Psi_0|\hat{T}_2\Psi_0} \longleftarrow \sum_{i<j}\sum_{a<b} t_{ij}^{ab} \cdot 
\bcontraction[1ex]{I^\dagger }{J}{^\dagger}{B}
\bcontraction[2ex]{}{I}{^\dagger J^\dagger B}{A}
\bcontraction[1ex]{I^\dagger J^\dagger BA a^\dagger }{b}{^\dagger }{j}
\bcontraction[2ex]{I^\dagger J^\dagger BA }{a}{^\dagger b^\dagger j}{i}
\contraction[1ex]{I^\dagger J^\dagger B}{A}{ a^\dagger }{b}
\contraction[2ex]{I^\dagger J^\dagger }{B}{A }{a}
\contraction[3ex]{I^\dagger }{J}{^\dagger BA a^\dagger b^\dagger }{j}
\contraction[4ex]{}{I}{^\dagger J^\dagger BA a^\dagger b^\dagger j}{i}
I^\dagger J^\dagger BA a^\dagger b^\dagger ji
\end{equation}
contributing to the CCSD doubles residual $R_{IJ}^{AB}$. In Table \ref{E2T2.tab}, the IRGs of zero, one and two (full) contractions in the occupied and virtual index spaces are shown. 

For no contractions at all, the IRGs simply reflect the index symmetry of the participating cluster and external super lines. For the given example it is $i<j$ and $I<J$ in the occupied space and $a<b$ and $A<B$ in the virtual space originating from the operators $\hat{E}_{AB}^{IJ}$ and $\hat{E}_{ij}^{ab}$, respectively. There are no active IRGs since there are no active indices involved. 

The first shown contractions involve the indices $I$ and $i$ (displayed in red) as well as $A$ and $b$ (displayed in blue). Each contraction merges two distinct nodes of the IRGs such that all edges connected to either of the nodes now connect to a single merged node. For the occupied indices, nodes $I$ and $i$ are merged to a single node $I=i$, which now connects to two outgoing edges to the nodes $J$ and $j$ representing the index relation $(I=i) < (J, j)$. 

\onecolumngrid
\begin{center}
\begin{table}
\caption{\label{vwxy.tab}Index relation graphs of uncontracted, singly and doubly contracted active indices $v$, $w$, $x$ and $y$. The indices are coupled through (e.g.) cluster super lines and possess the relations $v\neq w$ and $x \neq y$.}
\begin{tabular}{c|c|c}
$\bcontraction[1.5ex]{v^\dagger}{\ldots}{ }{w}
\bcontraction[2.5ex]{}{v}{^\dagger\ldots w}{\ldots}
\bcontraction[1.5ex]{v^\dagger\ldots w\ldots \quad\quad\quad\quad x^\dagger}{\ldots}{ }{y}
\bcontraction[2.5ex]{v^\dagger\ldots w\ldots \quad\quad\quad\quad }{x}{^\dagger\ldots y}{\ldots}
v^\dagger\ldots w\ldots \quad\quad\quad\quad x^\dagger\ldots y\ldots$
& 
$\contraction[1.5ex]{}{\highlight[myred]{v}}{^\dagger\ldots w\ldots \quad\quad\quad\quad x^\dagger\ldots }{\highlight[myred]{y}}
\bcontraction[1.5ex]{\highlight[myred]{v}^\dagger}{\ldots}{ }{w}
\bcontraction[2.5ex]{}{\highlight[myred]{v}}{^\dagger\ldots w}{\ldots}
\bcontraction[1.5ex]{\highlight[myred]{v}^\dagger\ldots w\ldots \quad\quad\quad\quad x^\dagger}{\ldots}{ }{\highlight[myred]{y}}
\bcontraction[2.5ex]{\highlight[myred]{v}^\dagger\ldots w\ldots \quad\quad\quad\quad }{x}{^\dagger\ldots \highlight[myred]{y}}{\ldots}
\highlight[myred]{v}^\dagger\ldots w\ldots \quad\quad\quad\quad x^\dagger\ldots \highlight[myred]{y}\ldots$ 
& 
$
\contraction[1.5ex]{v^\dagger\ldots }{\highlight[myblue]{w}}{\ldots \quad\quad\quad\quad }{\highlight[myblue]{x}}
\contraction[3ex]{}{v}{^\dagger\ldots \highlight[myblue]{w}\ldots \quad\quad\quad\quad \highlight[myblue]{x}^\dagger\ldots }{y}
\bcontraction[1.5ex]{v^\dagger}{\ldots}{ }{\highlight[myblue]{w}}
\bcontraction[2.5ex]{}{v}{^\dagger\ldots \highlight[myblue]{w}}{\ldots}
\bcontraction[1.5ex]{v^\dagger\ldots \highlight[myblue]{w}\ldots \quad\quad\quad\quad \highlight[myblue]{x}^\dagger}{\ldots}{ }{y}
\bcontraction[2.5ex]{v^\dagger\ldots \highlight[myblue]{w}\ldots \quad\quad\quad\quad }{\highlight[myblue]{x}}{^\dagger\ldots y}{\ldots}
v^\dagger\ldots \highlight[myblue]{w}\ldots \quad\quad\quad\quad \highlight[myblue]{x}^\dagger\ldots y\ldots$ 
\\[15pt] \hline
\rule{0pt}{3.2cm}\cincludegraphics[scale=1]{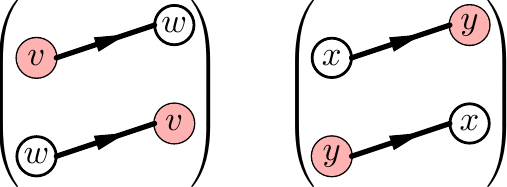}  &  \cincludegraphics[scale=1]{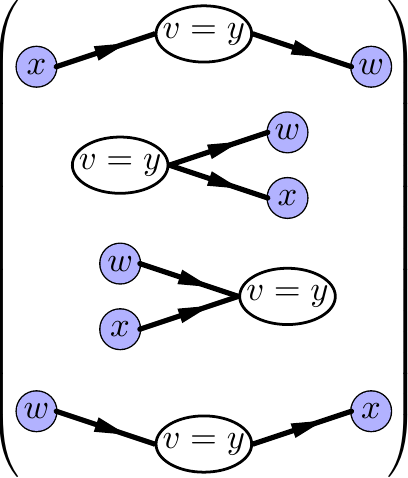}  &  \cincludegraphics[scale=1]{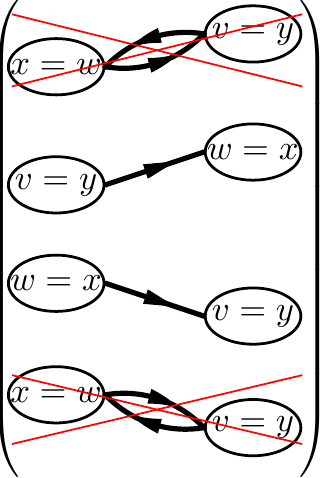}
\end{tabular} 
\end{table}
\end{center}
\twocolumngrid

For the virtual indices, on the other hand, the nodes $A$ and $b$ are unified. The outgoing and incoming lines of the nodes $A$ and $b$, respectively, are connected to a unified node $A=b$ such that the index relation $a < (A=b) < B$ is formed.

Finally, the contractions of $J$ and $j$ (displayed in green) and $B$ and $a$ (displayed in gray) complete the full contraction pattern. For the occupied indices, the two outgoing lines from node $I=i$ to nodes $J$ and $j$ are unified resulting in the final index relation $(I=i) < (J=j)$. For the virtual indices, nodes $B$ and $a$ are merged resulting in a looped diagram. Clearly, this represents an impossible index relation $(B = a) < (A = b) < (B = a)$ such that the whole term vanishes. 

A special case originating from the presence of $\neq$ relations, emerges in the active orbital space $\mathbb{A}$. Whenever such a relation is encountered, multiple IRGs are required to represent the index relations for $\mathbb{A}$ alone. Consider, e.g., the full contraction excerpt (assuming the two active contractions are on different spin paths) 
\begin{equation*}
\bcontraction[1ex]{\ldots v^\dagger}{\ldots }{}{w}
\bcontraction[2ex]{\ldots }{v}{^\dagger\ldots w}{\ldots}
\bcontraction[1ex]{\ldots v^\dagger\ldots w\ldots x^\dagger}{\ldots}{ }{y}
\bcontraction[2ex]{\ldots v^\dagger\ldots w\ldots }{x}{^\dagger\ldots y}{\ldots}
\contraction[1ex]{\ldots v^\dagger\ldots }{w}{\ldots }{x}
\contraction[2ex]{\ldots}{v}{^\dagger\ldots w\ldots x^\dagger\ldots }{y}
\ldots v^\dagger\ldots w\ldots x^\dagger\ldots y\ldots\,,
\end{equation*}
\\
where indices $v$ and $w$ as well as $x$ and $y$ are coupled via $v \neq w$ and $x \neq y$, respectively. In Table \ref{vwxy.tab}, the uncontracted, singly and doubly contracted IRGs of the given excerpt are shown.

For the uncontracted case, the unaltered index relations of the two cluster super lines are obtained. In this particular case, however, two graphs for each super line are possible. This is because the relation $v\neq w$ may be expanded to $v<w \lor w<v$ representing two IRGs. Exactly the same holds for the indices $x$ and $y$ connected by $x<y \lor y<x$. To graphically represent these choices, all valid IRGs are gathered in brackets such that the full index relation requires all possible choices. 

The first contraction (highlighted in red) merges indices $v$ and $y$. Since the indices are from two separate cluster super lines with two IRG choices each, a total of four merged IRGs is possible. These include the two linear chains $x < (v=y) < w$ and $w < (v=y) < x$ as well as the two v-shape relations $(v=y) < (w,x)$ and $(w,x) < (v=y)$. 

The second contraction (highlighted in blue) combines indices $x$ and $w$, which are now on the same IRGs. Through these unions, the two linear chain relations are rendered invalid due to occurring loops. The two v-shape relations, however, remain intact. From the original four IRG possibilities, only two survive as the final index relation $(v=y) < (w=x) \lor (w=x) < (v=y)$.

\subsection{Generation and implementation of final equations}
\label{FinalEquations}

To arrive at a final CC implementation of the generated Goldstone diagrams, the latter need to be translated to code of a given programming language. To achieve an implementation of somewhat manageable computational cost, it is mandatory to factorize the generated equations. Here, we apply a local, i.e. term/diagram-wise, factorization of each Goldstone diagram. While this is far from the global optimum, it still yields optimal scaling results in $\mathbb{O}$, $\mathbb{A}$ and $\mathbb{V}$. 

One major complication when evaluating the optimal factorization order of a given tensor product is the estimation of computational cost for different contraction routes. This is particularly difficult for routes including non-trivial index relations. As it turns out, the IRGs defined in the previous subsection \ref{IndexRelations} can easily be used to estimate this complexity, i.e. the number of iterations, of the represented index-iterating loops. 

\begin{center}
\begin{figure}[h]
\includegraphics[scale=1]{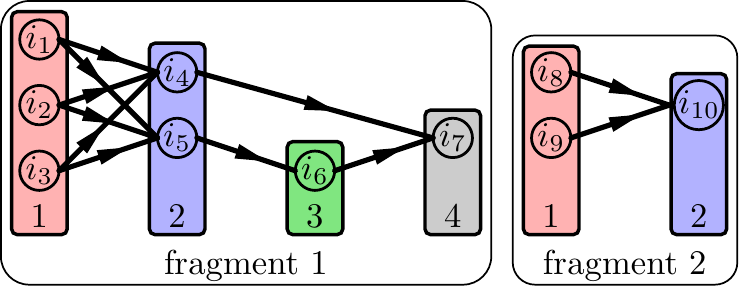}
\caption{\label{hierarchy_example.fig}Decomposition of exemplary index relation graph into disconnected fragments and hierarchy levels.}
\end{figure}
\end{center}

Each IRG may be decomposed into (i) its disconnected fragments and (ii) its different hierarchy levels. While decomposition (i) merely splits unconnected index relations, decomposition (ii) sorts the related indices according to their dependency from each other. Disconnected fragments scale independently of each other such that their complexities may be multiplied. Consider, e.g., the exemplary IRG in Figure \ref{hierarchy_example.fig} relating the indices $i_1$ to $i_{10}$.

Clearly, there are two disconnected fragments. The overall complexity of the IRG is given by the product of the complexities of its disconnected fragments 1 and 2. Within each fragment, the indices group to different hierarchy levels 1 -- 4 for fragment 1 and 1 -- 2 for fragment 2, respectively. A simple approach to sort indices into hierarchy levels is to iteratively extract nodes without incoming lines until no nodes are left. The extracted nodes during each iteration then represent the different hierarchy levels. This procedure, of course, requires the absence of any loops in the IRG. 

Trivially, any linear chain IRG composed of singly filled hierarchy levels only, possesses a complexity of $\binom{N}{m}$ for $N$ elements of the given index space and $m$ IRG nodes.

Each additional node, attaches a singly connected, i.e. $<$ related, index to the chain thereby increasing its complexity from $\binom{N}{m}$ to $\binom{N}{m+1}$. 

Unfortunately, valid IRGs may appear in completely different shapes and sizes as seen by the previous example. Whenever hierarchies containing multiple index nodes are encountered, a complexity evaluation must take all possible relations of these indices to each other into account. Consider, e.g., fragment 2 of the IRG shown in Figure \ref{hierarchy_example.fig}. Here, the two indices $i_8$ and $i_9$ are categorized to hierarchy level 1, i.e. have no index relation to each other. Therefore, the complexity of fragment 2 must consider all possible relations of $i_8$ and $i_9$ given by
\begin{equation*}
(i_8 = i_9) < i_{10} \quad \lor \quad i_8 < i_9 < i_{10}\quad\lor\quad i_9 < i_8 < i_{10}\,.
\end{equation*}

These in turn represent new (linearized) IRGs with trivially assignable complexities of $\binom{N}{2}$, $\binom{N}{3}$ and $\binom{N}{3}$, respectively. The overall complexity of fragment 2 is therefore given by  
\begin{equation*}
O\left((i_8, i_9) < i_{10} \right) = \binom{N}{2} + 2\binom{N}{3}\,.
\end{equation*}

In this work, arbitrary IRGs are analyzed with respect to their complexity using a recursive algorithm. This algorithm (in pseudocode) is listed as Algorithm \ref{recursive_complexity.alg}.

\begin{algorithm}[h]
\caption{\label{recursive_complexity.alg}Recursive complexity evaluation algorithm (in pseudocode) for disconnected fragments of arbitrary index relation graphs.}
\SetAlgoLined
\DontPrintSemicolon
\LinesNumbered
\KwIn{disconnected IRG fragment \texttt{f}, size of orbital space \texttt{N}}
\KwData{complexity \texttt{c}}
decompose \texttt{f} into hierarchy levels \texttt{h}\;
\If{all levels \texttt{h} are composed of single elements}
{
  \texttt{c += }binomial(\texttt{N}, no. of elements in \texttt{h})\;
  \Return{}\tcc*[r]{recursion end} 
}

find first degenerate hierarchy \texttt{d} in \texttt{h}\;
\For{all possible relations of the nodes in \texttt{d}}
{
  \texttt{f\_copy} = \texttt{f}\;
  merge identical indices in \texttt{f\_copy}\;
  add new relational edges to \texttt{f\_copy}\;
  recursive call for \texttt{(f\_copy, N)}\;
}
\end{algorithm}

In the presented algorithm, an arbitrary fragment \texttt{f}, the size of the corresponding orbital space \texttt{N} and the bookkeeping variable \texttt{c} for the calculated complexity are used. In line 1, the given fragment \texttt{f} is decomposed into its hierarchy levels \texttt{h} (as shown in figure \ref{hierarchy_example.fig}). If all hierarchy levels are singly populated, the given IRG fragment corresponds to a linear chain with a trivially assignable complexity. In terms of the algorithm, this constitutes the recursion end. The binomial complexity is added to \texttt{c} (line 3) and the call is returned (line 4). 

In case of at least one multiple-populated, i.e. degenerate, hierarchy level \texttt{d}, the main recursion loop (lines 7-12) is entered. Here, all possible index relations of the degenerate index nodes are enforced. In this work, we evaluate all possible index relations as all unique combinations of $=$ and $<$ connections between the different nodes. This is because, both $=$ and $<$ possess straight forward IRG representations, which may be applied to copies of \texttt{f}. For each combination of index relations among the degenerate nodes, a separate copy of \texttt{f} is created (line 8). In these copies, all identical nodes, i.e. nodes with a $=$ connection, are merged (line 9) while edges are added to all nodes connected by a $<$ relation (line 10). This leaves the altered copy \texttt{f\_copy}, where the degeneracy \texttt{d} was removed.  

To demonstrate the functionality of the algorithm, the full recursion tree of the exemplary IRG fragment given in Figure \ref{hierarchy_example.fig} as fragment 1 is shown in Figure \ref{RecursionExample.fig}. Given the initial IRG fragment, degenerate nodes are encountered in hierarchy levels 1 and 2 highlighted in red and blue, respectively. In the first recursion level, the degeneracy of the red nodes is removed. In total, there are 13 different $=$ and/or $<$ relations possible. These include 

\begin{itemize}
\item one possibility for three identical nodes ($i = j = k$), 

\onecolumngrid
\begin{center}
\begin{figure}[h]
\includegraphics[angle=-90, width=\textwidth]{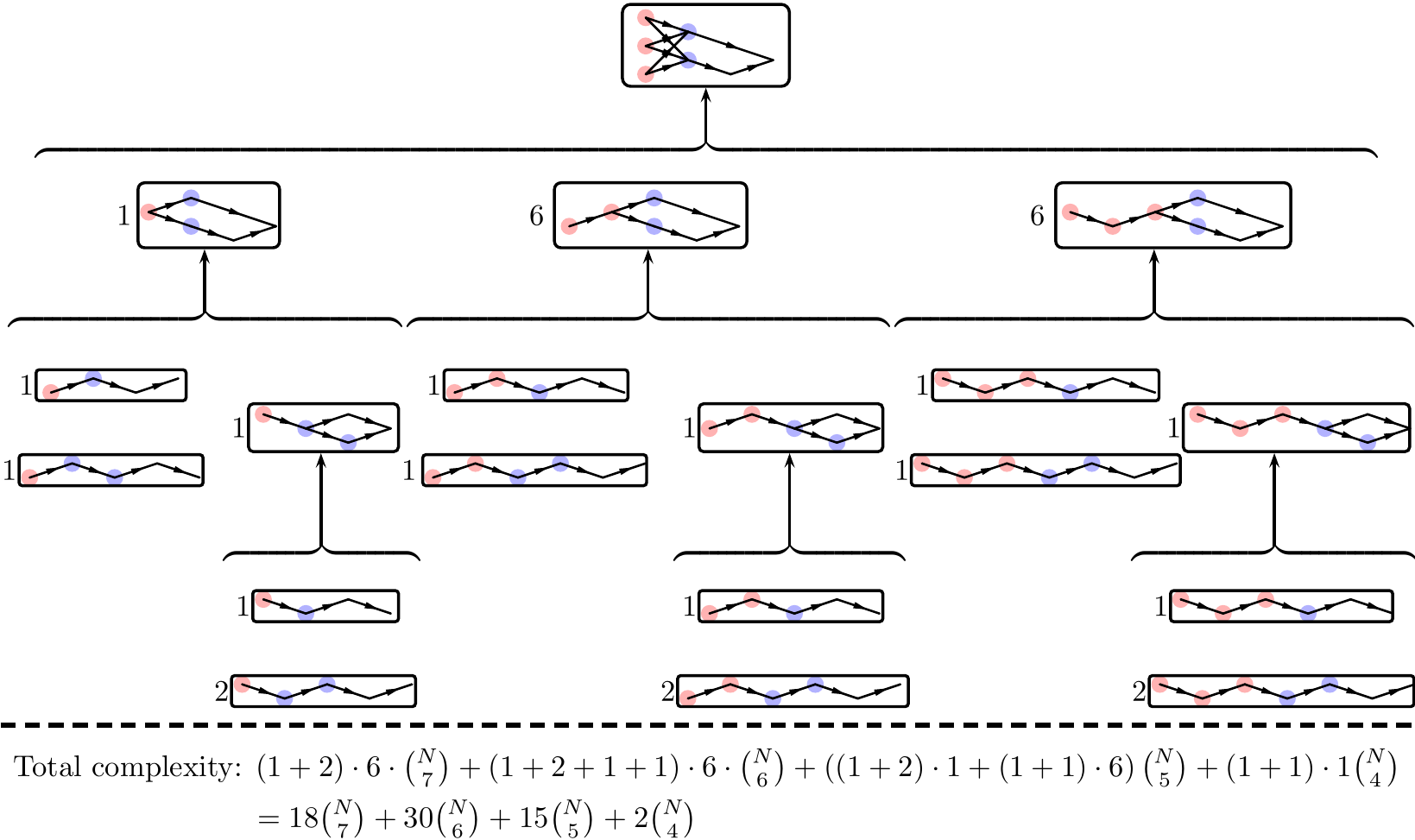}
\caption{\label{RecursionExample.fig}Recursion tree of the exemplary index relation graph given in Figure \ref{hierarchy_example.fig} containing two degenerate hierarchy levels (highlighted by red and blue nodes). The node labels as well as irrelevant edges are hidden to increase readability.}
\end{figure}
\end{center}
\twocolumngrid

\item six possibilities for two identical nodes ($(i = j) < k$, $k < (i = j)$, $(i = k) < j$, $j < (i = k)$, $(j = k) < i$ and $i < (j = k)$) and 
\item six possibilities for no identical nodes ($i < j < k$, $i < k < j$, $j < i < k$, $j < k < i$, $k < i < j$ and $k < j < i$).  
\end{itemize} 

Since all the red nodes are equally connected to the next hierarchy level (the blue nodes), the particular ordering of the nodes (in $<$ relations) is not important -- it leads to identical complexities. Different graphs only emerge from the different amounts of identical nodes. This is because
\begin{itemize}
\item[(1)] two nodes connected by two identical lines (no loop) represent the same IRG as two nodes connected by a single line:
\begin{center}
\includegraphics[scale=1]{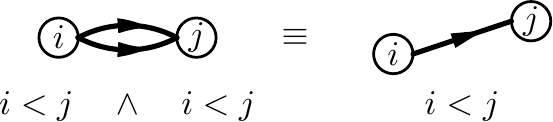}
\end{center}
\item[(2)] lines that skip certain hierarchy levels are irrelevant due to transitivity:
\begin{center}
\includegraphics[scale=1]{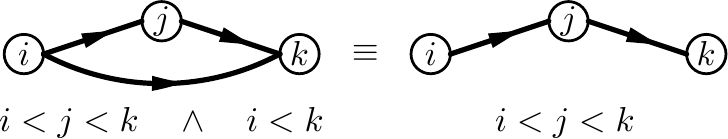}
\end{center}
\end{itemize}
Both criteria (1) and (2) are applied to the graphs of Figure \ref{RecursionExample.fig} such that doubled as well as irrelevant lines are neglected. 

For each of the 13 generated graphs with removed degeneracies in the red hierarchy level, the algorithm proceeds to the degeneracies of the blue nodes representing the next hierarchy level. There are three possibilities for the two blue nodes, which include 
\begin{itemize}
\item one possibility for two identical nodes ($i = j$) and 
\item two possibilities for no identical nodes ($i < j$ and $j < i$).
\end{itemize}

In contrast to the red nodes of hierarchy level 1, the ordering of the $<$ relation of the blue nodes matters. If the upper blue node is placed before the lower blue node, a linear chain IRG is obtained. If the upper blue node is placed after the lower blue node, however, a new degenerate hierarchy is formed. This new degenerate hierarchy needs to be eliminated in a third step of the recursion tree leading to three linear chain IRGs. For these, the ordering of the $<$ relations again is of no concern. 

Finally, the total complexity of the initial IRG fragment may be calculated by the addition of all complexities of the linear chain IRGs of the lowest recursion tree branches, i.e. the recursion endings as shown at the bottom of Figure \ref{RecursionExample.fig}. Please note that IRGs of identical complexities originating from arbitrary $<$ orderings are collected into groups with an assigned magnitude in Figure \ref{RecursionExample.fig}. These need to be included multiplicatively when evaluating the total complexity.

In this work, the presented complexity evaluation (c.f. Algorithm \ref{recursive_complexity.alg}) is used to optimize contraction routes of tensor products. First, each generated Goldstone diagram is translated to its algebraic form. Then all possible tensor products in all possible combinations are evaluated with respect to their complexity. This is done by extracting the participating indices of the tensor product from the full IRG of the diagram. This extracted IRG may then be analyzed w.r.t. its complexity using Algorithm \ref{recursive_complexity.alg}. Once all total complexities of all factorization routes per diagram are evaluated, the optimal route is determined. This is done by comparing the calculated complexities for exemplary orbital spaces of given numbers $N_O$, $N_A$ and $N_V$ and choosing the cheapest route. 

The optimal routes are then automatically translated to code either adding to $E_{\text{corr}}$ or to a residual tensor entry. The generated code is called by a separate CC equation solver, which employs the recently published Newton-Krylov method\cite{Yang_2020} to accelerate the residual convergence.

\newpage
\section{Application}
\label{Application}

In this section, the presented spin-adapted and spin-complete (SASC) CCSD implementation is applied to several small molecular test systems. The factorization routes of all CC equations was optimized w.r.t. the examplary orbital space sizes $N_{\mathbb{O}} = 6$, $N_{\mathbb{A}} = 3$ and $N_{\mathbb{V}} = 18$.

The effects of spin adaption and/or spin completeness in the non-orthogonal and orthogonal CC framework were already investigated in greater detail in our previous work\cite{Herrmann2021}. Here, we merely present some proof-of-concept applications to show the feasibility of the presented implementation. 

In subsection \ref{BCHConvergence}, the convergence of the correlation energy w.r.t. the BCH series truncation is investigated. Through several test calculations, a reasonable truncation at quadruply nested commutators is motivated. Subsection \ref{SpinorbitalComparison} features a comparison of SASC-CCSD correlation energies to (spin-contaminated) spin orbital CCSD and CCSD(T). 

\subsection{Convergence of BCH series truncation}
\label{BCHConvergence}

Due to active indices in the substitution operators of the cluster operator (c.f. equations \ref{Singles.eq} and \ref{Doubles.eq}), non-vanishing $\contraction{}{\hat{T}}{}{\hat{T}}\hat{T}\hat{T}$ and $\contraction{}{\hat{T}}{}{\hat{H}}\hat{T}\hat{H}_N$ contractions are possible. These lead to a non-terminating BCH series expansion in general. In our previous work\cite{Herrmann2021}, we investigated atomic systems H-N in the 6-31G basis set\cite{Hehre1972} with different orthogonal and non-orthogonal spin-adapted CC methods. We found that in 

\onecolumngrid
\begin{center}
\begin{table}[h] 
\caption{\label{BCHConvergence.tab} Correlation energies and Euclidean amplitude norms for SASC-CCS and SASC-CCSD for different atomic and molecular systems in the cc-pVDZ basis set\cite{Dunning1989} for a BCH series truncation of orders 1--4. The geometries of BeH and OH were taken from experimantal data\cite{Huber1979}. ROHF calculations were conducted using the PySCF\cite{Sun2015, Sun2018} program package with convergence thresholds of $10^{-14}$\,a.u. in both energy and density. CC residuals were converged to $10^{-7}$\,a.u..}
\begin{tabular}{c|c|c||c|c||c|c|c}

\textbf{Molecule} & $S = S_z$ & \textbf{BCH} & $E_{\text{corr}}(\text{CCS})$ & \textbf{Norm} $\hat{T}_1$ & $E_{\text{corr}}(\text{CCSD})$ & \textbf{Norm} $\hat{T}_1$ & \textbf{Norm} $\hat{T}_2$ \\ \hline 
\multirow{4}{*}{N} & \multirow{4}{*}{$\frac{3}{2}$} & 
   1 & $-0.020\,797\,5$ & $0.094\,110\,0$ & $-0.092\,428\,7$ & $0.086\,446\,7$ & $0.146\,818\,4$ \\ 
&& 2 & $-0.021\,020\,4$ & $0.093\,448\,8$ & $-0.091\,043\,8$ & $0.084\,021\,0$ & $0.143\,558\,9$ \\ 
&& 3 & $-0.021\,016\,9$ & $0.093\,432\,7$ & $-0.091\,035\,8$ & $0.084\,003\,4$ & $0.143\,539\,8$ \\
&& 4 & $-0.021\,016\,9$ & $0.093\,432\,7$ & $-0.091\,035\,9$ & $0.084\,003\,4$ & $0.143\,539\,9$ \\ \hline 
\multirow{4}{*}{BeH} & \multirow{4}{*}{$\frac{1}{2}$} & 
   1 & $-0.000\,777\,8$ & $0.040\,681\,7$ & $-0.040\,148\,2$ & $0.072\,134\,4$ & $0.137\,898\,8$ \\ 
&& 2 & $-0.000\,864\,1$ & $0.040\,548\,1$ & $-0.039\,172\,8$ & $0.058\,511\,2$ & $0.128\,839\,5$ \\
&& 3 & $-0.000\,864\,3$ & $0.040\,535\,7$ & $-0.039\,177\,6$ & $0.058\,507\,7$ & $0.128\,848\,9$ \\ 
&& 4 & $-0.000\,864\,3$ & $0.040\,535\,6$ & $-0.039\,177\,5$ & $0.058\,507\,3$ & $0.128\,848\,5$ \\ \hline
\multirow{4}{*}{OH} & \multirow{4}{*}{$\frac{1}{2}$} & 
   1 & $-0.009\,762\,0$ & $0.059\,997\,0$ & $-0.172\,095\,1$ & $0.064\,071\,0$ & $0.139\,366\,5$ \\ 
&& 2 & $-0.009\,989\,8$ & $0.059\,769\,8$ & $-0.169\,502\,6$ & $0.059\,346\,1$ & $0.136\,227\,5$ \\
&& 3 & $-0.009\,988\,9$ & $0.059\,762\,1$ & $-0.169\,503\,9$ & $0.059\,346\,9$ & $0.136\,231\,6$ \\ 
&& 4 & $-0.009\,988\,9$ & $0.059\,762\,1$ & $-0.169\,503\,9$ & $0.059\,346\,9$ & $0.136\,231\,6$
\end{tabular}
\end{table} 
\end{center}

\begin{center}
\begin{table}[h]
\caption{\label{SpinorbitalComparison.tab}Collection of CC correlation energies for molecular systems in different high-spin states using the cc-pVDZ basis set\cite{Dunning1989}. The given correlation energies include SASC-CCS and SASC-CCSD as presented in this work and spin orbital CCSD and CCSD(T) from the PySCF\cite{Sun2015, Sun2018} program package. For all CC calculations, the percentage of the recovered correlation energy w.r.t. the spin orbital CCSD(T) calculation are given. All geometries were assembled from experimental data. Geometries of BeH, CH, NH, OH, N$_2$ were assembled from \cite{Huber1979}, BH$_2$ from \cite{Herzberg1966} and CH$_2$ from \cite{Kuchitsu1998}. The ground state geometry was used for all calculations. ROHF calculations were also conducted using PySCF employing convergence criteria of $10^{-10}$\,a.u. in both energy and density. The CC residuals were converged to $10^{-7}$\,a.u. for all calculations.}
\begin{center} 
\begin{tabular}{c|c||c||c|c|c|c||c|c|c}
\centering
\multirow{2}{*}{\textbf{Molecule}} & \multirow{2}{*}{\textbf{$S=S_z$}} & \multirow{2}{*}{\textbf{$E_{\text{ROHF}}$}} & \multicolumn{4}{c||}{\textbf{$E_{\text{corr}}$ (SASC CC)}} & \multicolumn{3}{c}{$E_{\text{corr}}$ (\textbf{PySCF CC})} \\
& & & \textbf{S} & \textbf{\% SD(T)} & \textbf{SD} &  \textbf{\% SD(T)} & \textbf{SD} & \textbf{\% SD(T)} & \textbf{SD(T)} \\
\hline
\multirow{3}{*}{BeH} & $1/2$ & $-15.149\,436\,3$ & $-0.000\,864\,3$ & 2.18\% & $-0.039\,177\,5$ & 98.98\% & $-0.039\,156\,8$ & 98.93\% & $-0.039\,581\,0$ \\ 
& $3/2$ & $-14.967\,063\,4$ & $-0.000\,040\,3$ & 0.44\% & $-0.009\,198\,1$ & 99.76\% & $-0.009\,197\,5$ & 99.75\% & $-0.009\,220\,1$ \\
& $5/2$ & $-10.782\,898\,7$ & $\phantom{-}0.000\,000\,0$ & 0.00\% & $-0.015\,600\,7$ & 99.47\% & $-0.015\,600\,7$ & 99.47\% & $-0.015\,683\,2$ \\
\hline

\multirow{4}{*}{CH} & $1/2$ & $-38.268\,780\,0$ & $-0.009\,529\,8$ & 8.47\% & $-0.110\,685\,4$ & 98.43\% & $-0.110\,538\,0$ & 98.30\% & $-0.112\,452\,1$\\ 
& $3/2$ & $-38.278\,581\,8$ & $-0.012\,514\,8$ & 15.28\% & $-0.081\,241\,3$ & 99.17\% & $-0.080\,817\,5$ & 98.65\% & $-0.081\,924\,6$ \\
& $5/2$ & $-37.823\,593\,8$ & $-0.000\,137\,2$ & 0.43\% & $-0.031\,592\,3$ & 99.55\% & $-0.031\,590\,7$ & 99.54\% & $-0.031\,736\,0$ \\
& $7/2$ & $-26.823\,173\,6$ & $\phantom{-}0.000\,000\,0$ & 0.00\% & $-0.019\,921\,8$ & 99.57\% & $-0.019\,921\,8$ &  99.57\% & $-0.020\,007\,2$\\
\hline

\multirow{5}{*}{NH} & $0$ & $-54.857\,694\,8$ & $\phantom{-}0.000\,000\,0$ & 0.00\% & $-0.153\,304\,1$ & 95.70\% & $-0.153\,304\,1$ & 95.70\% & $-0.160\,185\,3$ \\ 
& $1$ & $-54.959\,577\,3$ & $-0.019\,020\,4$ & 14.25\% & $-0.132\,160\,8$ & 99.02\% & $-0.131\,866\,2$ & 98.80\% & $-0.133\,469\,2$ \\
& $2$ & $-54.663\,862\,3$ & $-0.020\,448\,1$ & 20.68\% & $-0.098\,106\,7$ & 99.23\% & $-0.097\,671\,3$ & 98.79\% & $-0.098\,869\,7$ \\
& $3$ & $-53.655\,520\,4$ & $-0.000\,211\,1$ & 0.68\%  & $-0.030\,775\,6$ & 99.59\% & $-0.030\,774\,2$ & 99.59\% & $-0.030\,901\,4$ \\
& $4$ & $-38.071\,856\,5$ & $\phantom{-}0.000\,000\,0$ & 0.00\% & $-0.018\,432\,1$ & 99.75\% & $-0.018\,432\,1$ & 99.75\% & $-0.018\,477\,8$ \\
\hline

\multirow{4}{*}{OH} & $1/2$ & $-75.390\,011\,3$ & $-0.009\,988\,9$ & 5.84\% & $-0.169\,503\,9$ & 99.07\% & $-0.169\,317\,6$ & 98.96\% & $-0.171\,100\,0$ \\ 
& $3/2$ & $-75.137\,787\,6$ & $-0.020\,481\,3$ & 14.74\% & $-0.137\,953\,2$ & 99.25\% & $-0.137\,626\,0$ & 99.01\% & $-0.138\,995\,5$ \\
& $7/2$ & $-72.737\,568\,1$ & $-0.000\,254\,9$ & 0.97\% & $-0.026\,192\,9$ & 99.70\% & $-0.026\,191\,8$ & 99.69\% & $-0.026\,272\,2$ \\
& $9/2$ & $-51.878\,407\,9$ & $\phantom{-}0.000\,000\,0$ & 0.00\% & $-0.018\,330\,0$ & 99.88\% & $-0.018\,330\,0$ & 99.88\% & $-0.018\,352\,4$ \\
\hline

\multirow{4}{*}{BH$_2$} & $1/2$ & $-25.751\,334\,5$ & $-0.003\,439\,3$ & 3.82\% & $-0.088\,862\,6$ & 98.82\% & $-0.088\,739\,8$ & 98.68\% & $-0.089\,926\,7$ \\ 
& $3/2$ & $-25.552\,158\,9$ & $-0.022\,558\,8$ & 25.00\% & $-0.089\,152\,4$ & 98.81\% & $-0.088\,266\,3$ & 97.82\% & $-0.090\,229\,9$ \\
& $5/2$ & $-25.183\,807\,8$ & $-0.000\,074\,9$ & 0.25\%  & $-0.029\,326\,7$ & 99.29\% & $-0.029\,325\,2$ & 99.29\% & $-0.029\,536\,0$ \\
& $7/2$ & $-18.021\,221\,3$ & $\phantom{-}0.000\,000\,0$ & 0.00\% & $-0.021\,532\,5$ & 99.31\% & $-0.021\,532\,5$ & 99.31\% & $-0.021\,681\,1$ \\
\hline

\multirow{3}{*}{CH$_2$} & $0$ & $-38.859\,108\,7$ & $\phantom{-}0.000\,000\,0$ & 0.00\% & $-0.138\,223\,7$ & 97.53\% & $-0.138\,223\,7$ & 97.53\% & $-0.141\,719\,3$ \\ 
& $1$ & $-38.921\,074\,5$ & $-0.013\,294\,1$ & 10.86\% & $-0.120\,937\,9$ & 98.81\% & $-0.120\,576\,8$ & 98.52\% & $-0.122\,389\,2$ \\
& $3$ & $-37.992\,164\,0$ & $-0.000\,146\,1$ & 0.40\%  & $-0.036\,432\,9$ & 99.39\% & $-0.036\,430\,9$ & 99.39\% & $-0.036\,655\,9$ \\
\hline  
\multirow{1}{*}{N$_2$} & $0$ & $-108.954\,139\,0$ & $\phantom{-}0.000\,000\,0$ & 0.00\% & $-0.313\,064\,3$ & 96.33\% & $-0.313\,064\,3$ & 96.33\% & $-0.324\,998\,2$ \\ 
\end{tabular}
\end{center}
\end{table}
\end{center}
\newpage
\twocolumngrid

most of the conducted calculations the correlation energy was already properly converged to 13 digits after the decimal point at a BCH truncation of quadruply nested commutators. For all cases, the maximal deviation was $10^{-12}$\,a.u. in the correlation energy w.r.t. a quintuply nested commutator truncation.

Therefore, in this work, we generated CCSD equations for up to quadruply nested commutators. To support this level of truncation, the convergence of $E_{\text{corr}}$ as well as the Euclidean norm of the $\hat{T}_1$ and $\hat{T}_2$ amplitudes with increasing BCH series truncation level was investigated. The results are collected in Table \ref{BCHConvergence.tab}.

Already for a BCH series truncation of order 3, all presented correlation energies for both CCS and CCSD are reasonably well converged. In particular, no noticeable change from a truncation of order 3 to order 4 was detected within the first six digits after the decimal point. In comparison to the correlation energy, the amplitude norms present a much more sensitive measure for the BCH convergence. Still, only negligible deviations in both $\hat{T}_1$ and $\hat{T}_2$ norms from BCH truncations of order 3 to order 4 were found. It is reasonable to assume that the same rapid convergence holds for comparable systems. 

\subsection{Comparison to spin orbital CC}
\label{SpinorbitalComparison}

Several test calculations of the presented SASC-CCSD were conducted to analyze its functionality. Initially, all calculations reported in our previous work\cite{Herrmann2021} (high-spin states of atomic hydrogen to nitrogen in the 6-31G basis set\cite{Hehre1972}), where a reference implementation of the SASC operators represented in the FCI basis was used, were reproduced. The correctness of the presented implementation was confirmed by comparing correlation energies and residuals for the converged amplitudes. Both were found in perfect agreement for up to 13 digits after the decimal point (residual mean squares were converged to $10^{-14}$\,a.u.). Therefore, it is reasonable to assume that the generated CC equations and factorizations are correct. 

Additionally, several test calculations for small molecular systems in the cc-pVDZ basis set\cite{Dunning1989} were performed. Table \ref{SpinorbitalComparison.tab} contains a comparison of SASC-CC to spin orbital CCSD and CCSD(T) correlation energies. In particular, the percentage of recovered correlation energy w.r.t. spin orbital CCSD(T) energies are shown.

While SASC-CCS recovered between 0.00\% and 25.00\%, SASC-CCSD recovered between 95.70\% and 99.88\% of the CCSD(T) correlation energies. In direct comparison, spin orbital CCSD results of all singlet ($S=0$) and all maximal high-spin results ($S=N/2$) are identical to the SASC-CCSD results. This is because in these cases, spin orbital CCSD is already properly spin-adapted and spin-complete. Hence, the equality of the results acts as another proof of concept for the presented implementation. 

In all other SASC-CCSD results, a small but noticeable improvement of the correlation energies compared to spin orbital CCSD is detected. These improvements reach from roughly $0.001$\,m$E_H$ to $0.886$\,m$E_H$ (0.01\% to 0.98\% w.r.t.\ spin orbital CCSD(T)) for octet OH and quartet BH$_2$, respectively. While correlation energy differences of up to $1$\,m$E_H$ might be significant when aiming for chemical accuracy, the effects of the SASC framework on the correlation energy seems to be negligible in comparison to the added complexity through the incorporation of spin-adapted operators into the CC framework for most calculations. Despite these findings, we still expect SASC-CC to produce superior wave functions when compared to spin orbital CC. These may be of particular importance when calculating spin-dependent molecular properties. A similar train of thought was also followed in our previous work\cite{Herrmann2020a, Herrmann2021} as well as e.g. by Datta and Gauss\cite{Datta2019}, who employ the spin-adapted CC framework to predict hyperfine coupling tensors.

\section{Conclusion}

A rigorously spin-adapted and spin-complete coupled cluster singles and doubles (CCSD) implementation capable of treating arbitrary high-spin open-shell states was reported. Following our previous work\cite{Herrmann2020a}, we employed the generated, L\"owdin-type substitution operators (c.f. equations \ref{Singles.eq} and \ref{Doubles.eq}) to ensure proper spin adaption as well as spin completeness of the CC wave function. While we focus on the CCSD truncation of the full CC operator for this work, the presented scheme is directly applicable to higher substitution ranks.

In section \ref{Methodology}, the generation of factorized CC equations was outlined. From the initial BCH expansion of the CCSD equations, Wick terms were generated using Wick's theorem. In subsection \ref{Wick}, we introduced a scheme to generate Wick terms on the basis of spatial rather than spin orbital indices. Through connecting lines below the second quantized operator string, we were able to evaluate prefactors arising from spin summations.

To remove redundancies of the generated Wick terms, we represented the latter by specially crafted spin-free Goldstone diagrams such that summarizable terms corresponded to isomorphic graphs (c.f. subsection \ref{Goldstone}). In order to do so, we added new diagrammatic features for active and identical lines to our Goldstone diagram definition. We then introduced cluster super lines to incorporate the special tensor symmetry required for the spin-adapted substitution operators. Finally, we added auxiliary vertices to our topologies to represent relations between the participating cluster operator indices. 

In subsection \ref{IndexRelations}, we introduced index relation graphs (IRGs) to evaluate, keep track and analyze index relations occurring in the tensor products of the CC equations.  These IRGs may then be used to estimate complexities of different tensor products (c.f. subsection \ref{FinalEquations}). Each single Goldstone diagram was translated to factorized \texttt{C++} code employing an optimal factorization route determined through complexity calculations of all possible factorization routes. The final codes were then compiled in batch-wise dynamically linkable library files to be called by a separate spin-adapted and spin-complete (SASC) CC driver program. 

The developed SASC-CCSD method was successfully applied to several small test molecules throughout section \ref{Application}. In subsection \ref{BCHConvergence}, the convergence of SASC-CCS and SASC-CCSD methods w.r.t. the BCH series truncation was investigated for three examples. Due to active indices in the cluster operator (representing both occupied and virtual orbitals in the spinorbital CC picture), the BCH series does not terminate after quadruply nested commutators in general. In all presented examples, however, proper convergence of the correlation energy as well as the Euclidean amplitude norms to an accuracy of at least $10^{-7}$\,a.u. was reported. These findings agree with our previous analysis for atomic test calculations\cite{Herrmann2021}. 

In subsection \ref{SpinorbitalComparison}, SASC-CCS and SASC-CCSD correlation energies were compared to spin orbital CCSD and CCSD(T) correlation energies. For all singlet ($S=0$) and maximum high-spin ($S=N/2$) calculations, SASC-CCSD reproduced the spin orbital CCSD results. Since spin orbital CCSD is properly spin-adapted and spin-complete in those cases, these findings were expected and act as a proof of concept for the presented implementation. In all other calculations, small improvements of the spin orbital CCSD correlation energies were found for SASC-CCSD. While the effects of spin adaption and spin completeness seem to be small for the correlation energy (as also discussed in e.g. \cite{Herrmann2020a, Herrmann2021, Datta2019}), they are expected to be significant for molecular properties, spin-dependent properties especially.



\end{document}